%% file: main.tex
\documentclass[twocolumn]{aastex631}

\usepackage{physics}

\newcommand*{\DLz}{$D_{\rm L}$--$z$\:}
\newcommand{\tnow}{t_{\rm now}}
\newcommand{\rhocheese}{\rho_{\rm cheese}}
\newcommand{\DL}{D_{\rm L}}
\newcommand{\ztarget}{z_{\rm target}}

\usepackage{soul}

\accepted{September 26, 2025}

\submitjournal{ApJ}

\shorttitle{$\DL$ Dispersion in Swiss-cheese Cosmology}
\shortauthors{Cheunchitra et al.}

\begin{document}

\title{Luminosity distance dispersion in Swiss-cheese cosmology as a function of the hole size distribution}

\correspondingauthor{T. Cheunchitra}
\email{tcheunchitra@student.unimelb.edu}

\author[0009-0001-2292-1914]{T. Cheunchitra}
\affiliation{School of Physics, University of Melbourne, Parkville, Victoria 3010, Australia}
\affiliation{OzGrav, Australian Research Council Centre of Excellence for Gravitational Wave Discovery, University of Melbourne, Parkville, Victoria 3010, Australia}

\author{A. Melatos}
\affiliation{School of Physics, University of Melbourne, Parkville, Victoria 3010, Australia}
\affiliation{OzGrav, Australian Research Council Centre of Excellence for Gravitational Wave Discovery, University of Melbourne, Parkville, Victoria 3010, Australia}

\author{R. L. Webster}
\affiliation{School of Physics, University of Melbourne, Parkville, Victoria 3010, Australia}



\input{abstract}




\input{section_1}

\input{section_2}

\input{section_3}
\input{section_4}

\input{section_5}
\input{section_6}
\input{section_7}
\input{section_8}



\input{thanks}

%


\software{scipy \citep{VirtanenEtAl2020}
          }



\appendix

\input{appendix_a}
\input{appendix_b}
\input{appendix_c}


\bibliography{swisscheese_bib}{}
\bibliographystyle{aasjournal}



\end{document}

%% file: abstract.tex
\begin{abstract}
    The luminosity distance-redshift ($\DL$--$z$) relation derived from Type Ia supernovae (SNe Ia) yields evidence for a nonzero cosmological constant.
    SNe Ia analyses typically fit to the functional form $\DL(z)$ derived theoretically from the homogeneous and isotropic Friedmann-Lemaitre-Robertson-Walker (FLRW) metric. 
    Yet, the metric in the epoch relevant to SNe Ia measurements deviates slightly from FLRW due to gravitational clumping of mass into large-scale structures like filaments and voids, whose sizes span many orders of magnitude. 
    The small deviation is modeled typically by scalar perturbations to the FLRW metric.
    Each line of sight to a SNe Ia passes through a random sequence of structures, so $\DL$ differs stochastically from one line of sight to the next. 
    Here, we calculate the $\DL$ dispersion in an exact Lemaitre-Tolman-Bondi Swiss-cheese universe with a power-law hole size distribution, as a function of the lower cut-off $R_{\rm min}$ and logarithmic slope $\gamma$.
    We find that the standard deviation of $\DL$ scales as $\sigma_{\DL} \propto z^{2.25\pm0.01} (R_{\rm min}/24\pm1\,{\rm Mpc})^{(0.157\pm0.003)\left[\gamma - (1.16\pm0.02)\right]}$ for redshifts in the range $0.5 \lesssim z \lesssim 2.1$.  
    The scaling shows that the $\DL$ dispersion is dominated by a few large voids rather than the many small voids.

\end{abstract}

%% file: section_1.tex
\section{Introduction}
\label{sec:intro}
    
    The matter-energy content of the universe can be inferred from the functional relationship between distances and redshifts of astrophysical objects, e.g.\ the luminosity distance-redshift (\DLz) relation of standard candles such as Type Ia supernovae (SNe Ia).
    The observed SNe Ia \DLz relation is approximately consistent with a nonzero cosmological constant $\Lambda$ (i.e.\ ``dark energy'') \citep{RiessEtAl1998, PerlmutterEtAl1999} and the widely-accepted $\Lambda$-cold-dark-matter ($\Lambda$CDM) concordance cosmology \citep{PeeblesRatra2003, PlanckCollaborationEtAl2020}.
    However, the concordance seems to be imperfect: there is some evidence of a mismatch between the Hubble constant $H_0$ implied by the SNe Ia \DLz relation and the $H_0$ value inferred from cosmic microwave background (CMB) measurements, known colloquially as the ``Hubble tension'' \citep{RiessEtAl2019, PlanckCollaborationEtAl2020}.

    Standard \DLz analyses infer cosmological parameters within the framework of the exactly homogeneous and isotropic Friedmann-Lemaitre-Robertson-Walker (FLRW) metric \citep{Friedmann1922, Robertson1935, Robertson1936, Robertson1936a, Walker1937}.  
    However, the late-time universe deviates from exact homogeneity and isotropy, because gravity aggregates matter into large-scale structures such as clusters and filaments interspersed by voids \citep{BolejkoEtAl2011, AluriEtAl2023}. 
    Voids with characteristic rectilinear extent $10 \lesssim L/(1 \, {\rm Mpc}) \lesssim 30$ are observed in galaxy surveys \citep{HoyleVogeley2004, PanEtAl2012} and cosmological simulations \citep{ContariniEtAl2022}.
 
    Inhomogeneity affects $\DL$ and $z$ in two ways.
    First, if the Milky Way occupies a special location in the universe, e.g.\ near the center of a large local void, the deviation from the FLRW metric perturbs $\DL$ and $z$ \citep{Celerier2000, AlexanderEtAl2009, RedlichEtAl2014, CamarenaEtAl2022, MazurenkoEtAl2024a, MazurenkoEtAl2024b}.    
    Second, when the first effect is small, deviations from FLRW are dominated by lensing by random sequences of clusters, filaments, and voids along different lines of sight, whose size distributions are determined by the physics of gravitational collapse and large-scale structure formation \citep{PressSchechter1974, ShethVanDeWeygaert2004, JenningsEtAl2013}. 
    Hence, $\DL$ differs stochastically from one line of sight to the next, unlike in a FLRW universe \citep{FlanaganEtAl2012, ClarksonEtAl2012b, FlanaganEtAl2013, Macpherson2023}. 
    Modelling the dispersion of $D_{\rm L}$ due to this stochasticity is the focus of this paper. Standard SNIa analyses model the $D_{\rm L}$ dispersion using a Gaussian distribution with redshift-dependent variance \citep[e.g.][]{ScolnicEtAl2018}. However, the $D_{\rm L}$ dispersion is known to be nongaussian, and attempts have been made to model its nongaussianity \citep{WangEtAl2002, HolzLinder2005, BolejkoFerreira2012, AlfradiqueEtAl2024}. More recent analyses aim to detect explicitly the sources of lensing and forward-model the $D_{\rm L}$ dispersion to improve parameter estimation \citep{AmendolaEtAl2010, KronborgEtAl2010, JonssonEtAl2010, SmithEtAl2013, ShahEtAl2024}.

    The dominant contribution to the $D_{\rm L}$ dispersion is the magnification of SNe Ia by clumps of matter, such as galaxy clusters \citep{FlanaganEtAl2012, FlanaganEtAl2013}; see Figure 6 in \citet{AlfradiqueEtAl2024} for example. In this paper, we focus instead on another contribution: the demagnification of SNe Ia due to voids. One fundamental question about the void contribution to the $D_{\rm L}$ dispersion is whether it is dominated by a few large voids or the many small voids.
    The outcome depends in part on $p(L)$, the probability density function (PDF) of the void size $L$. 
    Ideally, one would answer the question by ray-tracing through a general-relativistic simulation of the large-scale structure, as it evolves. 
    \cite{Macpherson2023} estimated the $\DL$ dispersion for a matter-only universe \citep{MacphersonEtAl2019} to be ${\sim}1\%$ at $z \sim 10^{-2}$.
    However, computational cost limits numerical resolution to ${\sim} 10$ Mpc, which may not be adequate to capture fully the aggregate effect of the smallest voids, e.g.\ if the low-$L$ end of $p(L)$ is steep enough. 
    An alternative approach is to calculate the $\DL$ dispersion analytically for exact solutions of the Einstein field equations \citep{VanderveldEtAl2008, FlanaganEtAl2012, FlanaganEtAl2013, FleuryEtAl2015a}, thereby avoiding resolution limitations at low $L$. 
    Exact solutions suited to this task include Swiss-cheese universes, where one models an inhomogeneous universe by embedding spherical underdensities (``holes'') in an otherwise homogeneous FLRW universe (``cheese'')\footnote{Historically, the adjective ``Swiss-cheese'' also refers to models in which the holes contain a central overdensity, representing clumps of masses like galaxies \citep{EinsteinStraus1945, Schuecking1954, Kantowski1969, FleuryEtAl2013, FleuryEtAl2013a, Fleury2014}. In this paper, we focus on voids, whose density is a minimum at the center.} \citep{VanderveldEtAl2008, FlanaganEtAl2012, FlanaganEtAl2013, FleuryEtAl2013, FleuryEtAl2013a, Fleury2014}. 
    The analytic form of the metric in a Swiss-cheese universe is known at all locations, so calculating $\DL$ and $z$ involves integrating the geodesic equations and Sachs optical equations \citep{Sachs1961}. 
    Swiss-cheese holes can be made arbitrarily smaller than the spatial resolution of numerical simulations.
    The low computational cost makes Monte Carlo experiments feasible \citep{HolzWald1998, VanderveldEtAl2008, FlanaganEtAl2012, FlanaganEtAl2013, FleuryEtAl2013a}.

    Previous efforts to calculate $\DL$ in a Swiss-cheese universe find ${\sim}1\%$ dispersion for $z\lesssim 1$ \citep{WambsganssEtAl1997, BrouzakisEtAl2008, FlanaganEtAl2012, FlanaganEtAl2013}.
    The latter references assume that all holes are the same size and larger than ${\sim} 30$ Mpc. 
    In this paper, we generalize previous calculations to consider $L \lesssim 30 \, {\rm Mpc}$ in order to determine analytically how $\DL$ dispersion depends on the low-$L$ cut-off of $p(L)$, denoted by $L_{\rm min}$, and its power-law exponent, denoted by $\gamma$. 
    The observational evidence for smaller holes is compelling: galaxy redshift surveys \citep{PellegriniEtAl1989, KauffmannFairall1991, PlionisBasilakos2002, HoyleVogeley2004, PanEtAl2012, NadathurHotchkiss2014, NadathurEtAl2015, ContariniEtAl2023, SongEtAl2025b} and cosmological simulations \citep{NadathurEtAl2015, ContariniEtAl2022, WilliamsEtAl2024} detect voids down to their resolution limit of ${\sim}$10 Mpc.
    Past work has indicated that it is sometimes sufficient to approximate an inhomogeneous cosmological metric by the FLRW metric with scalar perturbations, such as when the peculiar velocity is small \citep{IshibashiWald2005,Acoleyen2008,FlanaganEtAl2012,FlanaganEtAl2013}. Such an approach is efficient computationally. In this paper, we elect to employ an exact, nonperturbative Swiss-cheese model to avoid truncating the perturbative expansion, and instead develop a method to calculate $D_{\rm L}$ in such a model with low computational cost.
    
    The paper is organized as follows.
    In Section \ref{sec:swisscheese_model}, we summarize the construction of the Swiss-cheese model from exact solutions of the Einstein field equations.
    In Section \ref{sec:propagation_lightbeams}, we write down the equations of motion of a light beam in a Swiss-cheese universe, and discuss how to solve them to calculate $\DL$ and $z$ for a given source. 
    In Section \ref{sec:measuring_DL_dispersion}, we integrate the equations of motion numerically to obtain $\DL$--$z$ data in one fiducial Swiss-cheese universe, i.e. for one illustrative choice of $L_{\rm min}$ and $\gamma$. 
    In Section \ref{sec:varying_hole_distribution}, we calculate how the $\DL$ dispersion depends on $L_{\rm min}$ and $\gamma$. 
    In Section \ref{sec:void_distribution}, we compare the power-law void size distribution used in this paper to observed and simulated void catalogs.
    In Section \ref{sec:highz}, we extend the analysis to redshifts beyond what are currently accessible through SNe Ia, in preparation for gravitational wave studies in the future. 
    We comment briefly on astrophysical implications, including the SNe Ia \DLz relation, in Section \ref{sec:conclusion}.

%% file: section_2.tex
\section{Swiss-cheese cosmology}
\label{sec:swisscheese_model}

    We restrict our attention to a cosmological model which contains $\Lambda \neq 0$ and pressureless matter.
    Spacetime in the Swiss-cheese holes is described by the Lemaitre-Tolman-Bondi (LTB) metric  \citep{Tolman1934, Bondi1947}, the spherically-symmetric nonstatic dust solution of the Einstein field equations.
    In order to maintain consistency with the FLRW cheese, the density profile of the LTB holes must satisfy a matching condition.
    We present in this section a short review of the LTB metric (Section \ref{subsec:LTB}), a useful parameterization for integrating numerically the geodesic equations (Section \ref{subsec:friedmannlikeparam}), and a hole density profile consistent with the matching condition (Section \ref{subsec:junction-compensation}).

\subsection{LTB metric}
\label{subsec:LTB}
 
    In spherical comoving coordinates $(t, r, \theta, \phi)$, the spacetime interval for the LTB metric is commonly written as \citep{BolejkoEtAl2009}
    \begin{equation}
        \dd s^2 = -\dd t^2 + \frac{{Y'(t, r)}^2}{1+2E(r)} \dd r^2 + Y(t, r)^2\dd\Omega^2~,
        \label{eq:LTBmetric}
    \end{equation}
    with $\dd \Omega^2 = \dd \theta^2 + \sin^2 \theta \dd \phi^2$. 
    Primes signify derivatives with respect to $r$, and $E(r)$ is an arbitrary function determined by initial conditions. 
    The evolution of $Y(t, r)$ is governed by 
    \begin{equation}
        \dot{Y}^2(t, r) = \frac{\Lambda Y(t, r)^2}{3} + 2E(r) + \frac{2S(r)}{Y(t, r)}~,
        \label{eq:EoMforY}
    \end{equation}
    where a dot signifies a derivative with respect to $t$, and $S(r)$ is the gravitational mass contained within radius $r$,
    \begin{equation}
        S(r) = 4\pi G \int_0^r \dd \tilde{r} ~ \rho(t, \tilde{r}) Y(t, \tilde{r})^2 Y'(t, \tilde{r}) ~,
        \label{eq:Sdefn}
    \end{equation}
    where $\rho(t, r)$ denotes the matter density. 

\subsection{Friedmann-like parameterization of the LTB metric}
\label{subsec:friedmannlikeparam}
    
    It is convenient in this application to rewrite the LTB metric in the following way, to assist with the matching recipe in Section \ref{subsec:junction-compensation}. 
    Upon defining $a(t, r) = Y(t, r)/r$, $k(r) = -2E(r)/r^2$, and $\mu(r) = S(r)/r^3$, equations \eqref{eq:LTBmetric}, \eqref{eq:EoMforY}, and \eqref{eq:Sdefn} become
    \begin{align}
        \dd s^2 =&\, -\dd t^2 + a(t, r)^2 \nonumber \\ 
        &\, \times \left\{ \left[1 + \frac{ra'(t, r)}{a(t, r)}\right]^2\frac{\dd r^2}{1 - k(r)r^2} + r^2 \dd\Omega^2 \right\}~, \label{eq:FriedmannLikeLTBmetric} 
    \end{align}
    \begin{align}
        \dot{a}(t,r)^2 = \frac{2\mu(r)}{a(t, r)} - k(r) + \frac{\Lambda a(t, r)^2}{3}, \label{eq:FriedmannLikeEoM} 
    \end{align}
    and
    \begin{align}
        \mu(r) =&\, \frac{4\pi G}{r^3} \nonumber \\
        &\, \times \int_0^{r} \dd \tilde{r} ~ \rho(t, \tilde{r}) \tilde{r}^2 a(t, \tilde{r})^2 \qty[a(t, \tilde{r}) + \tilde{r}a'(t, \tilde{r})], \label{eq:FriedmannLikeMu}
    \end{align}
    respectively.
    We note that $a(t, r)$, $k(r)$, and $\mu(r)$ do not diverge as $r$ approaches zero, when there is no point mass or curvature singularity at $r=0$ \citep{BolejkoEtAl2009}.
    We choose the notation of the arbitrary functions in this metric suggestively; the LTB metric reduces to the FLRW metric, when the dust density is homogeneous, and $a(t, r)$ and $k(r)$ reduce to the FLRW scale factor and spatial curvature respectively.
    Integrating equation \eqref{eq:FriedmannLikeEoM} yields
    \begin{equation}
        \int^{a(t, r)}_{0} \dd \tilde{a} \; \qty[\frac{2\mu(r)}{\tilde{a}} - k(r) + \frac{\Lambda \tilde{a}^2}{3}]^{-1/2} = t - t_{\rm B}(r)~,  \label{eq:integratedFriedmannLikeEoM}
    \end{equation}
    in which the constant of integration $t_{\rm B}(r)$, colloquially called the bang time function, is a function of $r$ \citep{Zibin2008, BolejkoEtAl2009}.     
    
    The LTB metric is completely specified by $\Lambda$, and three out of four interdependent functions: $k(r)$, $\mu(r)$, $t_{\rm B}(r)$, and the gauge choice of the radial coordinate $r$.
    In this paper, we specify the LTB metric in the Swiss-cheese holes through the following recipe:
    \begin{enumerate}
        \item the radial coordinate is chosen such that the scale factor at present time $\tnow$ is $a(\tnow, r) = 1$;
        \item the bang time satisfies $t_{\rm B}(r) = 0$ at all locations; and
        \item the density profile of the holes at $\tnow$, $\rho(\tnow, r)$, is chosen such that the expansion of the hole is consistent with that of the cheese (see Section \ref{subsec:junction-compensation}). 
    \end{enumerate}
    We then calculate $k(r)$ by solving equation \eqref{eq:integratedFriedmannLikeEoM}, evaluated at time $\tnow$. 
    Setting the bang time function $t_{\rm B}(r)$ to be constant in the above recipe is a physical (i.e.\ not a gauge) choice. It means that there are only growing modes (no decaying modes) in the evolution of the density fluctuations; see Appendix A of \citet{Zibin2008} as well as \citet{HellabyKrasinski2006}. Constant bang time is a common choice for Swiss-cheese models \citep{BiswasNotari2008, BrouzakisEtAl2007}, but other choices have been made too \citep{MarraEtAl2007, MarraEtAl2008, VanderveldEtAl2008}.
    
\subsection{Embedding the holes in the cheese: mass compensation}
\label{subsec:junction-compensation}

    When constructing a Swiss-cheese model, one needs to ensure that the metrics match consistently across the boundary between the LTB holes and the FLRW cheese. 
    In this context, matching means that the metrics join smoothly, satisfying the Israel junction conditions \citep{Israel1966}.
    The first Israel condition is that the induced metric on both sides of the boundary is equal. 
    The second Israel condition is specified by equation (5) in \citet{KhakshourniaMansouri2001}, including the general case where there is a thin mass shell on the boundary between the LTB and FLRW regions. 
    In this paper, we restrict our attention to the case where there is no thin mass shell. 
    
    It is easy to see how the Israel conditions can be satisfied using the Friedmann-like parameterization.
    Consider the metric near a LTB hole with radius $R$.
    One can choose a coordinate system such that the metric is of the form \eqref{eq:FriedmannLikeLTBmetric}--\eqref{eq:FriedmannLikeMu} for $r < R$ (in the hole) as well as $r > R$ (in the cheese), with the additional requirements $k(r) = k_{\rm cheese}$ and $a(t, r) = a_{\rm cheese}(t)$ independent of $r$ in the cheese.
    In Appendix \ref{appx:junctioncondition}, we show that the Israel conditions reduce to three requirements: $a(t_{\rm now}, R) = a_{\rm cheese}(t_{\rm now})$, $k(R) = k_{\rm cheese}$, and $\mu(R) = 4\pi G \rhocheese/3$, where $\rhocheese$ is the matter density in the FLRW cheese. 
    The last requirement is commonly referred to as ``mass compensation'' \citep{VanderveldEtAl2008, FlanaganEtAl2012, FlanaganEtAl2013}.
    One can satisfy the above requirements by choosing a functional form of the matter density profile of the holes in equation \eqref{eq:FriedmannLikeMu} evaluated at $\tnow$. 
    That is, one chooses the form of $\Delta(r, R) = \rho(\tnow, r)/\rho_{\rm cheese}$. 
    
    One commonly-used parameterisation for the density profile of a cosmic void is the Hamaus-Sutter-Wandelt (HSW) profile, which fits well to stacked density profiles of voids detected in N-body simulations \citep{HamausEtAl2014}.
    For a spherical void of comoving radius $R$, the HSW profile is
    \begin{align}
        \Delta_{\rm HSW}(r, R) = 1 - \frac{\delta\qty[1 - (r/R_{\rm s})^{\alpha}]}{1 + (r/R_{\rm eff})^{\beta}} ~,
    \end{align}
    which is characterised by five dimensionless parameters, $\delta, \alpha, \beta$, and the ratios $R_{\rm s}/R$ and $R_{\rm eff}/R$.
    The HSW profile satisfies the mass compensation condition for specific combinations of these five parameters, which must be calculated numerically. 
    In this paper, we use instead a simpler one-parameter approximation,
    \begin{align}
        \Delta(r, R) = 1 - \delta \qty(1 - \frac{r^2}{3 c^2 R^2}) \exp \qty(-\frac{r^2}{2 c^2 R^2})~. \label{eq:Emmentaler}
    \end{align}
    In \eqref{eq:Emmentaler}, $c$ is a dimensionless parameter, and $\delta$ is the fractional underdensity at the center of the hole.  
    For ease of reference, we call the density profile \eqref{eq:Emmentaler} the Emmentaler density profile\footnote{After the protected designation of origin of Swiss-cheese.}. 
    The Emmentaler profile is a fiducial model chosen for computational purposes, and is not intended to match accurately to the models of voids inferred from observations \citep{HamausEtAl2014, StopyraEtAl2024}.
    
\begin{figure}[ht!]
        \centering
        \includegraphics[width=0.98\linewidth]{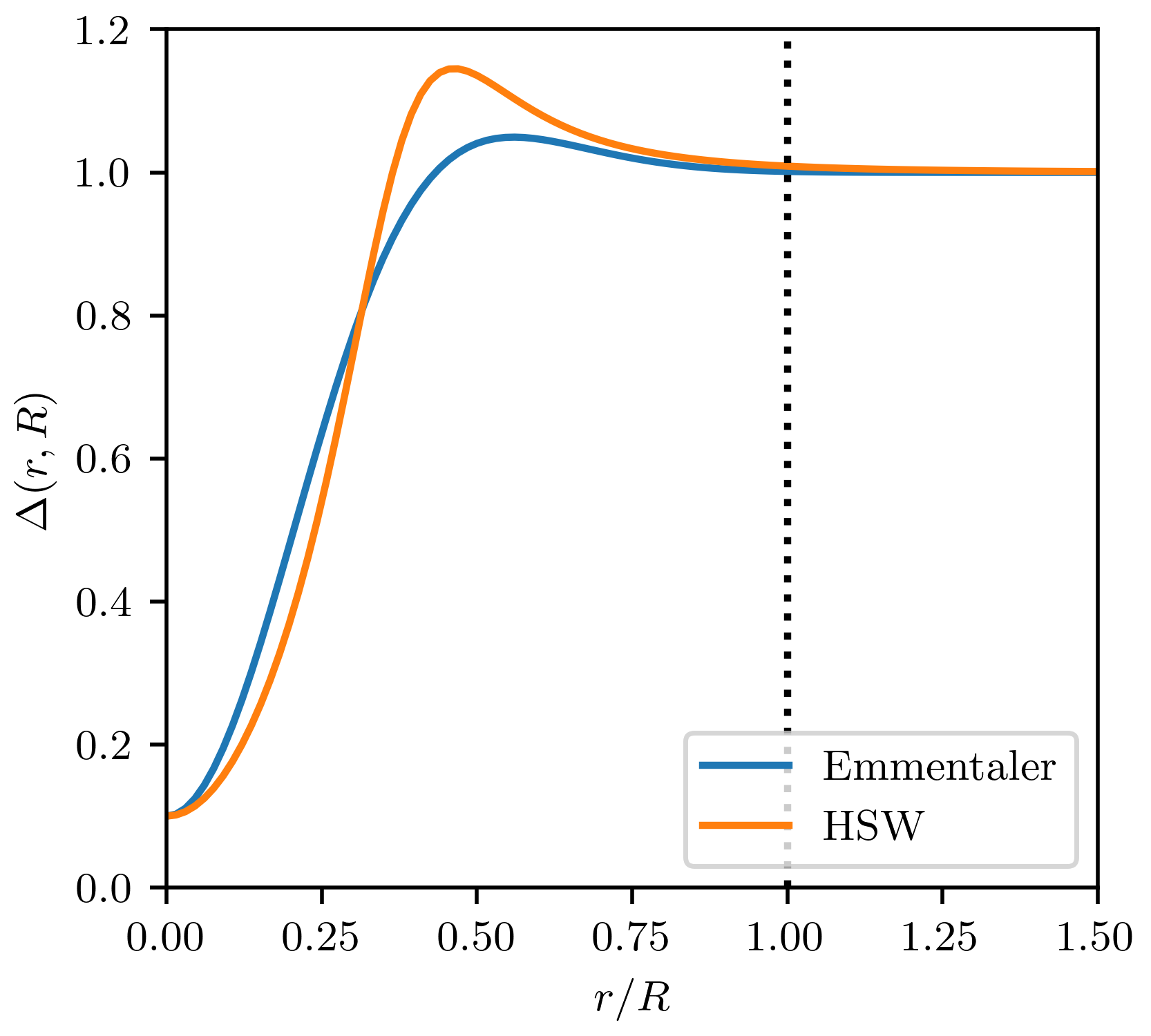}
        \caption{Comparison between the one-parameter Emmentaler density profile used in this paper (blue curve) and the five-parameter HSW density profile (orange curve). The Emmentaler profile is plotted with $c=0.25$, and the HSW density profile is plotted with $\delta=0.90$, $R_{\rm s}=0.36R$, $R_{\rm eff}=0.91R$ $\alpha=2.0$, $\beta=7.1$, which is approximately mass-compensated. The vertical dotted line marks $r=R$, where light beams exit one LTB region and enter another (see Section \ref{subsec:continuity_conditions}).}
        \label{fig:densityprofiles}
    \end{figure}

    The Emmentaler and HSW density profiles are compared in Fig.\ \ref{fig:densityprofiles}. 
    They look similar by design; both profiles exhibits a broad central underdensity surrounded by a sharp overdensity.
    Both profiles approaches unity rapidly as $r \rightarrow \infty$.
    Likewise, by substituting \eqref{eq:Emmentaler} into \eqref{eq:FriedmannLikeMu}, one can show that $\mu(r)$ for the Emmentaler profile approaches $4\pi G \rhocheese/3$ exponentially. 
    That is, the Emmentaler holes satisfy the mass compensation condition approximately at $r=R$. 
    The precision of this approximation is determined by the dimensionless parameter $c$, matching to better than $0.1 \%$ for $c \leq 0.25$. 
    One may be wary of this inexactness and how it affects the propagation of light beams at the hole-to-cheese boundary. 
    In this paper, however, we arrange the continuity conditions (see Section \ref{subsec:continuity_conditions}) so that the light beam does not encounter discontinuities in the density, as long as $c$ is the same for all holes. 

    The matching conditions mean that the metric in a Swiss-cheese hole of radius $R$ is completely determined by $\rho_{\rm cheese}$, $k_{\rm cheese}$, $\Lambda$, $\delta$ and $c$.
    These five quantities are held constant for all Swiss-cheese universes considered in this paper. 
    We define for convenience a reference density
    \begin{align}
        \rho_{\rm c} = 8.53 \times 10^{-27} \, \text{kg\,m}^{-3}~.
        \label{eq:rho_c}
    \end{align}
    This value is a fiducial number which is taken arbitrarily to be equal to the critical density inferred from the \textit{Planck} 2018 cosmic microwave background measurements \citep{PlanckCollaborationEtAl2020}. 
    The Swiss-cheese quantities are defined with respect to $\rho_{\rm c}$ in Table \ref{tab:SwissCheeseParams}. 
    
    \begin{table}[ht!]
        \centering
        \begin{tabular}{ll}
            \hline 
            Quantity                & Value \\
            \hline
            $\rho_{\rm cheese}$     & $0.3\rho_{\rm c}$  \\
            $k_{\rm cheese}$        & $0$ \\
            $\Lambda$               & $0.7 (8\pi G \rho_{\rm c})$ \\
            $c$                     & 0.25 \\
            $\delta$                & 0.9 \\
            \hline
        \end{tabular}
        \caption{Physical quantities specifying the metric in a Swiss-cheese hole.}
        \label{tab:SwissCheeseParams}
    \end{table}

%% file: section_3.tex
\section{Propagation of light beams}
\label{sec:propagation_lightbeams}

    The propagation of a light beam within the Swiss-cheese universe in Section \ref{sec:swisscheese_model} has two aspects. 
    The motion of the centroid of the beam cross-section is governed by the geodesic equations.
    The evolution of the cross-section is governed by the Sachs optical equations \citep{Sachs1961}. 
    These equations form a system of ordinary differential equations, which we present in Section \ref{subsec:EoMofBeam} in a modified form, which accelerates their integration.
    We derive Lorentz-invariant continuity conditions for joining light beams exiting one hole and entering another in Section \ref{subsec:continuity_conditions}, and describe how to obtain $\DL$ and $z$ from the integrated equations in Section \ref{subsec:extracting_observables}.

\subsection{Equations of motion in a single LTB hole}
\label{subsec:EoMofBeam}
    
    Without loss of generality, we orient the coordinates $(t, r, \theta, \phi)$ such that the light beam propagates in the plane $\theta=\pi/2$. 
    The geodesic equations can be written as \citep{MarraEtAl2007, Szybka2011}
    \begin{align}
        \dv{p^t}{s} =&\, - \frac{(\dot{a} + r\dot{a}')(a + ra')}{1 - kr^2} (p^r)^2 - \dot{a} a r^2 (p^{\phi})^2~, \label{eq:dpt_ds}\\
        \dv{p^r}{s} =&\, -\frac{2(\dot{a}+r\dot{a}')}{a+ra'}p^t p^r + \frac{1 - kr^2}{a + ra'} ra (p^\phi)^2 \nonumber \\
        &\, -\left[\frac{kr + k'r^2/2}{1-kr^2} + \frac{2a' + ra''}{a + ra'}\right] (p^r)^2~, \label{eq:dpr_ds}\\
        \dv{p^\phi}{s} =&\, -\frac{2\dot{a}}{a} p^t p^\phi - \frac{2(a+ra')}{ar} p^r p^\phi~, \label{eq:dpphi_ds} \\
        \dv{x^t}{s} =&\, p^t~, \label{eq:dxt_ds} \\
        \dv{x^r}{s} =&\, p^r~, \label{eq:dxr_ds} \\
        \dv{x^\phi}{s} =&\, p^\phi~, \label{eq:dxphi_ds} 
    \end{align}
    where $s$ is the affine parameter, $x^{\mu}$(s) is the trajectory of the centroid of the beam cross-section, and $p^{\mu}$ is defined through equations \eqref{eq:dxt_ds}--\eqref{eq:dxphi_ds}.
    In equations \eqref{eq:dpt_ds}--\eqref{eq:dpphi_ds}, we omit the functional dependence in $a[x^t(s), x^r(s)]$ and $k[x^r(s)]$ for clarity. 

    To calculate $\DL$, we need to know how the beam cross-section (de)focuses as it propagates. 
    Information about the expansion, shear, and twist of the beam cross-section is encoded in the Sachs optical equations \citep{Sachs1961}\footnote{We assume in this paper that the typical width of a beam cross-section is small relative to the characteristic rectilinear size of voids \citep{FleuryEtAl2013, Fleury2014}.}. 
    \citet{BrouzakisEtAl2008} and \citet{Szybka2011} demonstrated that the shear is small, when $\rho(t, r)$ is smooth.
    The twist cannot grow, if the beam is not initially twisted.
    Hence we neglect the effects of shear and twist in this paper.
    The remaining Sachs optical equation is the one describing the evolution of the expansion scalar, $\theta$, viz.
    \begin{equation}
         \dv{\theta}{s} = -\frac{1}{2}\theta^2 - R_{\alpha\beta}p^\alpha p^\beta ~, \label{eq:expansion_reduced}
    \end{equation}
    where $R_{\alpha\beta}$ is the Ricci tensor of the LTB metric. 
    The expansion scalar is related to the characteristic length scale of the beam cross-section $b$ via \citep{Sachs1961, Poisson2004, MisnerEtAl1973}
    \begin{equation}
        \theta = \frac{1}{b^2}\dv{b^2}{s}~. \label{eq:expansion_area}
    \end{equation}
    Substituting \eqref{eq:expansion_area} into \eqref{eq:expansion_reduced}, with
    \begin{align}
        \dv{b}{s} = J~, \label{eq:db_ds}
    \end{align}
    the Sachs optical equation for the expansion scalar becomes
    \begin{align}
        \dv{J}{s} = - R_{\alpha\beta}p^\alpha p^\beta b~. \label{eq:dJ_ds}
    \end{align}
    
    Equations \eqref{eq:dpt_ds}--\eqref{eq:dxphi_ds}, \eqref{eq:db_ds}, and \eqref{eq:dJ_ds} make up a system of ordinary differential equations which describes the propagation of a light beam in the Swiss-cheese universe.
    The state vector $\vb{V} = (p^t, p^r, p^\phi, x^t, x^r, x^\phi, b, J)$ contains the information one needs to calculate $\DL$ and $z$.
    In principle, one can integrate \eqref{eq:dpt_ds}--\eqref{eq:dxphi_ds}, \eqref{eq:db_ds}, and \eqref{eq:dJ_ds} numerically from the source to the observer, evaluating $a[x^t(s), x^r(s)]$, $k[x^r(s)]$, and their derivatives by solving \eqref{eq:integratedFriedmannLikeEoM} numerically at every $s$-step.
    In practice, however, solving \eqref{eq:integratedFriedmannLikeEoM} numerically is expensive computationally. 
    To overcome this issue, we include $a(s) = a[x^t(s), x^r(s)]$ and $k(s) = k[x^r(s)]$ in $\vb{V} = \qty(p^t, p^r, x^t, x^r, \phi, b, J, a, k)$.
    The associated equations of motion are
    \begin{align}
        \dv{a}{s} =&~ \dot{a} p^t + a' p^r~, \label{eq:da_ds} \\
        \dv{k}{s} =&~ k' p^r ~. \label{eq:dk_ds}
    \end{align}
    Full details of the integration algorithm are given in Appendix \ref{appx:metricfunctions}.

\subsection{Hole-to-hole continuity}
\label{subsec:continuity_conditions}

\begin{figure}[t]
        \centering
        \includegraphics[width=0.98\linewidth]{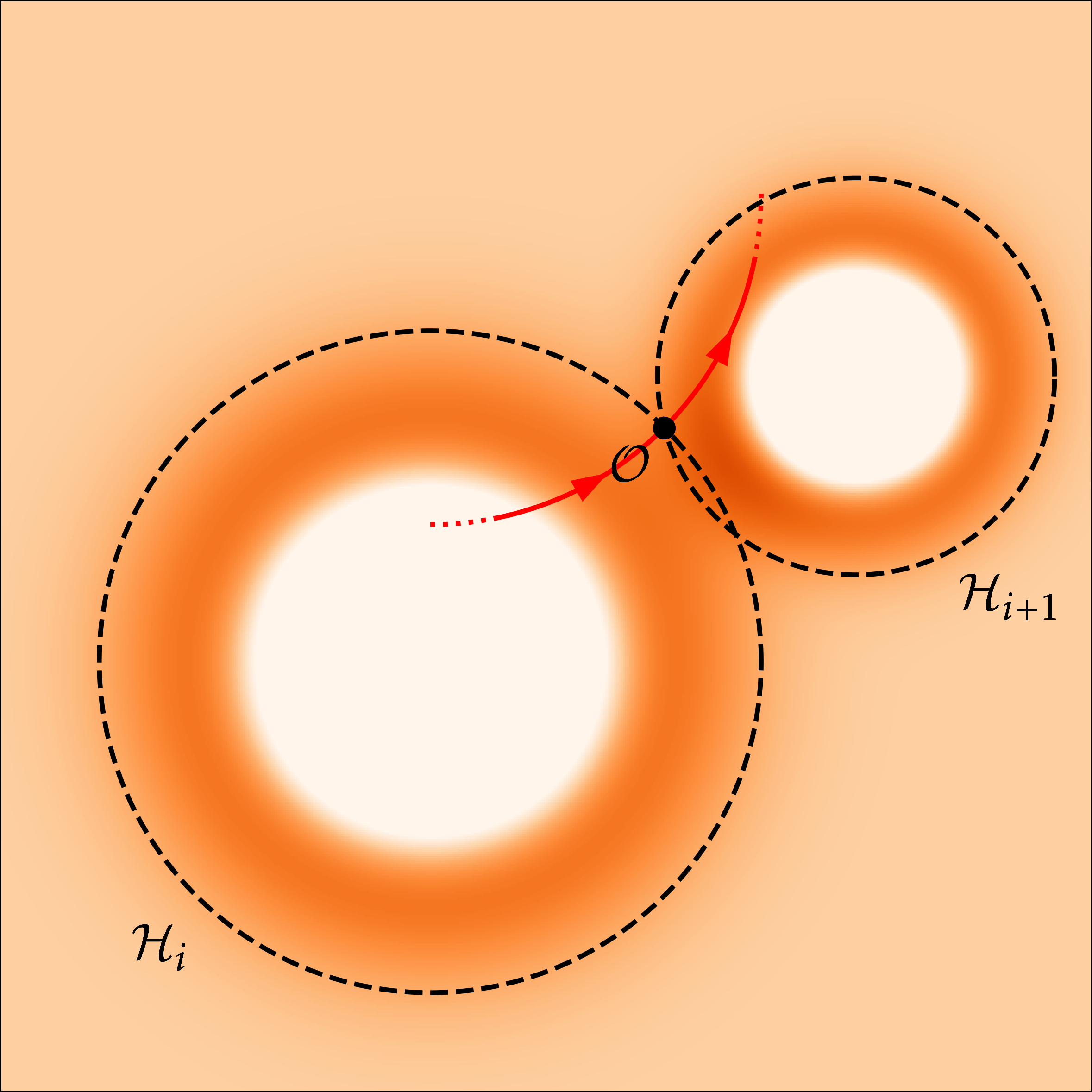}
        \caption{Schematic diagram showing continuity conditions between two adjacent Swiss-cheese holes. Darker shading corresponds to higher matter density. The geodesic (red arrow) exits hole $\mathcal{H}_{i}$ and enters hole $\mathcal{H}_{i+1}$ at point $\mathcal{O}$ (black dot), where the continuity conditions in Section \ref{subsec:continuity_conditions} are applied. The holes overlap slightly by design.}
        \label{fig:continuity}
    \end{figure}

    Equations \eqref{eq:dpt_ds}--\eqref{eq:dxphi_ds} and \eqref{eq:db_ds}--\eqref{eq:dk_ds} describe the propagation of a light beam through a single Swiss-cheese hole. 
    In general, the beam propagates through alternating regions of hole and cheese.
    One needs a prescription to turn the state vector of a light beam exiting one hole into the state vector of a light beam entering the cheese, and vice versa. 
    In practice, though, one does not need to consider propagation in the FLRW cheese separately from propagation in the LTB hole.
    With the density profile \eqref{eq:Emmentaler}, the FLRW cheese can be included in the spherical region described by the LTB metric by tuning $c$, such that there is a substantial region $R_{\rm cheese} < r < R$, where the density is close to $\rho_{\rm cheese}$.  
    With such tuning, the light beam propagates through both hole and cheese within a single LTB region described by \eqref{eq:dpt_ds}--\eqref{eq:dk_ds} and can then move onto the next LTB region without loss of generality.
    This setup is illustrated schematically in Fig.\ \ref{fig:continuity}.
    In this paper, we choose $c=0.25$.
    With this choice of $c$, the density of the Emmentaler profile at $r=R$ (marked with the dotted line in Fig.\ \ref{fig:densityprofiles}), where the light beam exits one hole and enters the next, is $\approx 0.1\%$ larger than $\rho_{\rm cheese}$. 
    The light beam propagates only a short distance in the portion of the LTB region that approximates the cheese.
    This setup matches qualitatively the observation that walls of cosmic voids are thin relative to their sizes \citep{WilliamsEtAl2024}.
    
    We derive the conditions for transforming $\vb{V}$ exiting one Swiss-cheese hole to $\vb{V}$ entering the next Swiss-cheese hole in Appendix \ref{appx:continuity}. 
    The approach is based on coordinate-invariant quantities, i.e.\ quantities observed by a fictional comoving observer at the contact point between the holes, which should be the same in the coordinate systems of both holes.  
    
\subsection{Extracting observables: \texorpdfstring{$\DL$}{D\_L} and \texorpdfstring{$z$}{z}}
\label{subsec:extracting_observables}

    Given the state vector at the point of emission (say, a SN Ia), denoted by $\vb{V}_{\rm e} = \qty(p^t_{\rm e}, p^r_{\rm e}, t_{\rm e}, r_{\rm e}, \phi_{\rm e}, k_{\rm e}, a_{\rm e}, b_{\rm e}, J_{\rm e})$, we can calculate the state vector at the point of observation (Earth), denoted by $\vb{V}_{\rm o} = \qty(p^t_{\rm o}, p^r_{\rm o}, t_{\rm o}, r_{\rm o}, \phi_{\rm o}, k_{\rm o}, a_{\rm o}, b_{\rm o}, J_{\rm o})$.
    In practice, we start with $\vb{V}_{\rm o}$ and integrate the equation of motion ``backwards'' to obtain $\vb{V}_{\rm e}$. 
    The Swiss-cheese universe is initialised at coordinate time $\tnow$, coinciding with the observation epoch, making it a natural starting point for the integration. 
    We then calculate $z$ and $\DL$ from the components of $\vb{V}_{\rm e}$.

    \subsubsection{Redshift}
        The energy of a photon with 4-momentum $\vb{p}$ observed by a comoving observer with 4-velocity $\vb{u}$ is given by $\mathcal{E} = - \vb{p} \cdot \vb{u}$. 
        In the comoving coordinates used for the LTB metric, the energy simplifies to $\mathcal{E} = -p^t$.
        The redshift $z$ of a photon is defined by
        \begin{equation}
            z = \frac{\mathcal{E}_{\rm e}}{\mathcal{E}_{\rm o}} - 1
            \label{eq:redshift_E}
        \end{equation}
        where $\mathcal{E}_{\rm e}$ and $\mathcal{E}_{\rm o}$ are the energies of the photon when it is emitted and observed respectively. 
        Assuming that the source and observer are comoving, we obtain
        \begin{equation}
            z = \frac{p^t_{\rm e}}{p^t_{\rm o}} - 1~.
            \label{eq:redshift_p}
        \end{equation}

    \subsubsection{Luminosity distance}

        \begin{figure}[t]
            \centering
            \includegraphics[width=0.98\linewidth]{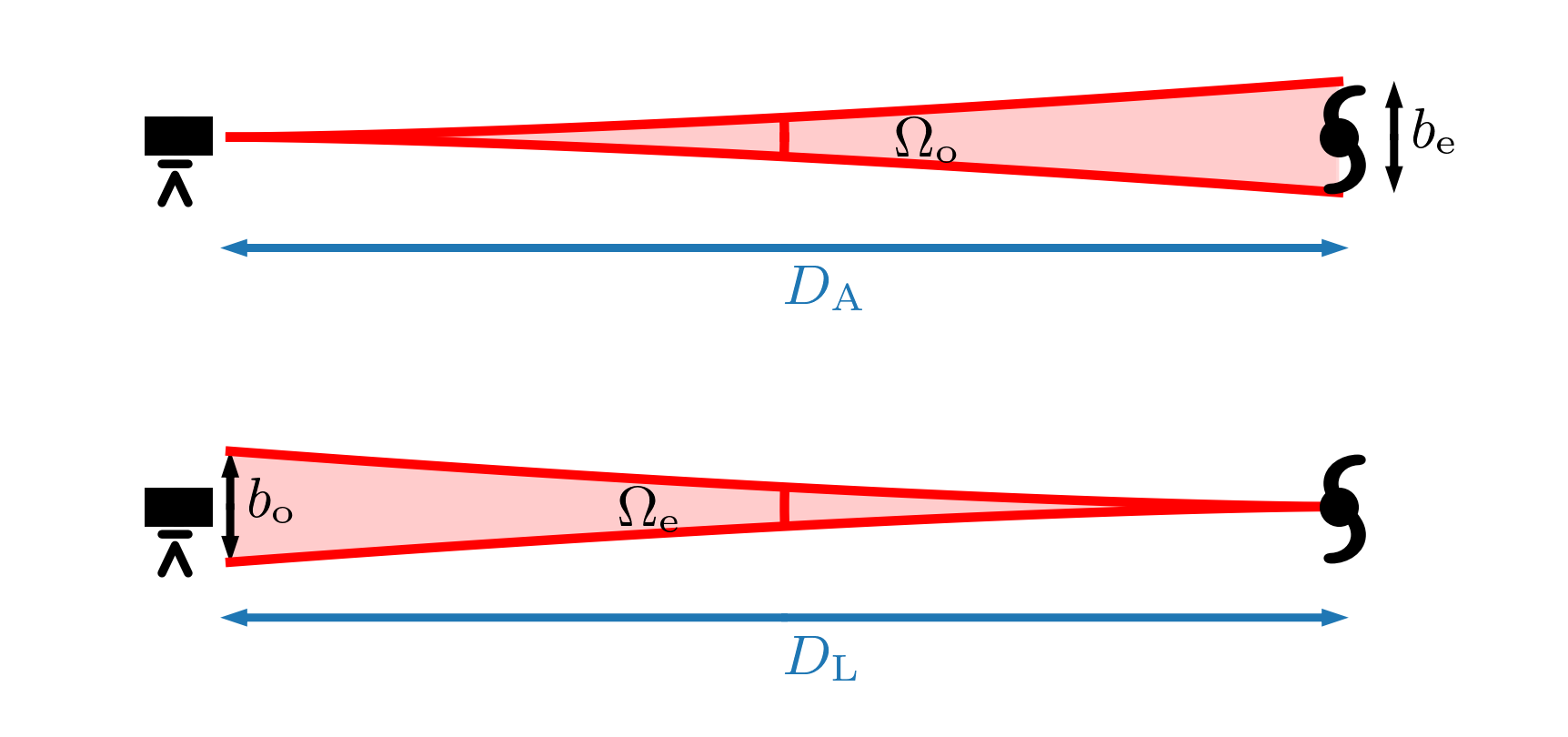}
            \caption{Schematic diagram of distance measures in cosmology of a source (galaxy) measured by the observer (telescope) and how they relate to the state variables of the beam. \textit{Top panel}: Angular diameter distance. \textit{Bottom panel}: Luminosity distance.}
            \label{fig:distances}
        \end{figure}
        
        A convenient way to calculate $\DL$ is through the angular diameter distance $D_{\rm A}$ via Etherington's reciprocity theorem \citep{Etherington1933}\footnote{In this paper, Etherington's reciprocity theorem is used as a calculational device. In practice, the observer measures directly the luminosity distance of the source.},
        \begin{align}
            D_{\rm A}(z) = (1+z)^{-2} \DL(z). \label{eq:Etherington}
        \end{align}
        The top panel of Fig.\ \ref{fig:distances} depicts the geometrical reciprocity. 
        The beam of light that defines the angular diameter of the source is initialized with $b_{\rm o}=0$.
        For a light beam which converges to a point, the quantity $J/p^t$ at the convergence point equals the square root of the solid angle subtended by the beam \citep{BrouzakisEtAl2007}, so we have $J_{\rm o} = \mathcal{E}_{\rm o}\Omega_{\rm o}^{1/2}$.
        
        The angular diameter distance inferred by the observer is the physical size of the object, divided by the apparent angular size, viz. $D_{\rm A} = b_{\rm e} \Omega_{\rm o}^{-1/2}$ and hence
        \begin{equation}
            \DL = (1+z)^2 b_{\rm e} \Omega_{\rm o}^{-1/2}~. \label{eq:luminositydistance}
        \end{equation}

%% file: section_4.tex
\section{Measuring the \texorpdfstring{$\DL$}{DL} dispersion at a given redshift}
\label{sec:measuring_DL_dispersion}

    Ideally, the $\DL$ dispersion at fixed redshift $z$ can be estimated by performing the following Monte Carlo experiment.
    Firstly, one integrates equations \eqref{eq:dpt_ds}--\eqref{eq:dxphi_ds} and \eqref{eq:db_ds}--\eqref{eq:dk_ds}, until the redshift \eqref{eq:redshift_p} reaches $z$.
    Secondly, one calculates $\DL$ using \eqref{eq:luminositydistance} from $\vb{V}_{\rm e}$ to obtain the datum $(\DL, z)$ for a single source.
    Repeating this recipe for several realizations of the Swiss-cheese universe gives a sample of $(z, \DL)$ data, from which one can deduce the $\DL$ dispersion at fixed $z$. 
    However, a source at fixed $z$ lies in the hole in some realizations of a Swiss-cheese universe and in the cheese in others. 
    For a source which lies in a hole, its beam cross-section is typically distorted more by the occupied hole than by the aggregate effect of intervening holes\footnote{Similar effects occur when putting observers in a hole instead of the cheese \citep{MarraEtAl2007, BolejkoEtAl2009}.}.
    In this paper, we concentrate on the effect of intervening holes, and place all sources in the cheese, i.e.\ at $r=R$ in the hole closest to the source.  
    It is reasonable to expect most SNe Ia to be in the overdense regions of the universe; \citet{PanEtAl2012} estimated that only $\sim 7\%$ of galaxies in the Sloan Digital Sky Survey Data Release 7 catalogue reside in voids.
    
    We generate the sample of $(z, \DL)$ around a fixed redshift $\ztarget$ using the following recipe: 

    \begin{enumerate}
        \item Initialise the light beam at the point of observation with photon energy $\mathcal{E}_{\rm o}$ and angular size of beam $\Omega_{\rm o}$. 

        \item Draw the cosine of the incident angle $\cos \alpha_{\rm o}$, where $\alpha_{\rm o}$ is the angle between the photon direction and the $r$-direction of the first hole, from a uniform distribution on the interval $[0, 1)$.
        
        \item Draw the radius $R$ of the first hole from a probability distribution $p(R)$. In this paper, we choose $p(R)$ to be a power law \citep{ShethVanDeWeygaert2004, ContariniEtAl2023, StopyraEtAl2024}, 
        \begin{equation}
            p(R) = \begin{cases}
                C R^{-\gamma} & R_{\rm min} \leq R \leq R_{\rm max} \\
                0 & \text{otherwise}
            \end{cases}~, \label{eq:powerlawPDF}
        \end{equation}
        where $C$ is a normalization constant, defined so that $\int_{ R_{\rm min}}^{R_{\rm max}} \, \dd R p(R) = 1$.
        This simple form of $p(R)$ parameterizes the size and abundance of the smallest holes through $R_{\rm min}$ and $\gamma$ respectively. 
        The lower cut-off $R_{\rm min}$ plays the role of $L_{\rm min}$ (see Section \ref{sec:intro}) in a spherical hole.

        \item Place the first hole so that the point of observation is at $r=R$, and $\phi=0$.

        \item Calculate the initial state vector $\vb{V}_{\rm o}$ from $\mathcal{E}_{\rm o}$, $\Omega_{\rm o}$, and $\cos\alpha_{\rm o}$, assuming $r_{\rm o} = R$. This calculation is the same as the transformation from the state vector to coordinate-invariant quantities derived in Appendix \ref{appx:continuity}. 
        
        \item Propagate the light beam through the Swiss-cheese hole by integrating equations \eqref{eq:dpt_ds}--\eqref{eq:dxphi_ds} and \eqref{eq:db_ds}--\eqref{eq:dk_ds} until $r=R$. 

        \item Use the continuity conditions described in Section \ref{subsec:continuity_conditions} and Appendix \ref{appx:continuity} to transform the final state vector of the first hole into the initial state vector of the second hole. As part of this, calculate the redshift $z'$ of the light beam at the exit point.

        \item Repeat steps 2 to 7 until $z' > \ztarget$ is achieved. 

        \item To ensure that the sample contains data with both $z < \ztarget$ and $z > \ztarget$, we set $\vb{V}_{\rm e}$ to be the state vector before the beam enters the final hole for half of the sample, and the state vector after the beam exits the final hole for the other half. 

        \item Calculate $z$ and $\DL$ by applying \eqref{eq:redshift_p} and \eqref{eq:luminositydistance} to $\vb{V}_{\rm e}$, and record the datum $(z, \DL)$ in the sample.
        
    \end{enumerate}
    The integration in step 4 is performed using the \textsc{LSODA} method\footnote{LSODA is a modified version of the Livermore Solver for Ordinary Differential Equations (LSODE). It detects stiffness of the problem at each step, and switches between the Adams method and the Backwards Differentiation Formula method automatically.} \citep{2019ascl.soft05021H, Petzold1983} implemented in the \textsc{scipy} package \citep{VirtanenEtAl2020}. 

    In the equations of motion, the units of 4-momentum components, spatial curvature, and solid angle cancel out. 
    These units are set arbitrarily by the choice of initial values, and do not affect the final outcomes (i.e.\ $\DL$ and $z$).
   
    \begin{figure}[ht!]
        \centering
        \includegraphics[width=0.98\linewidth]{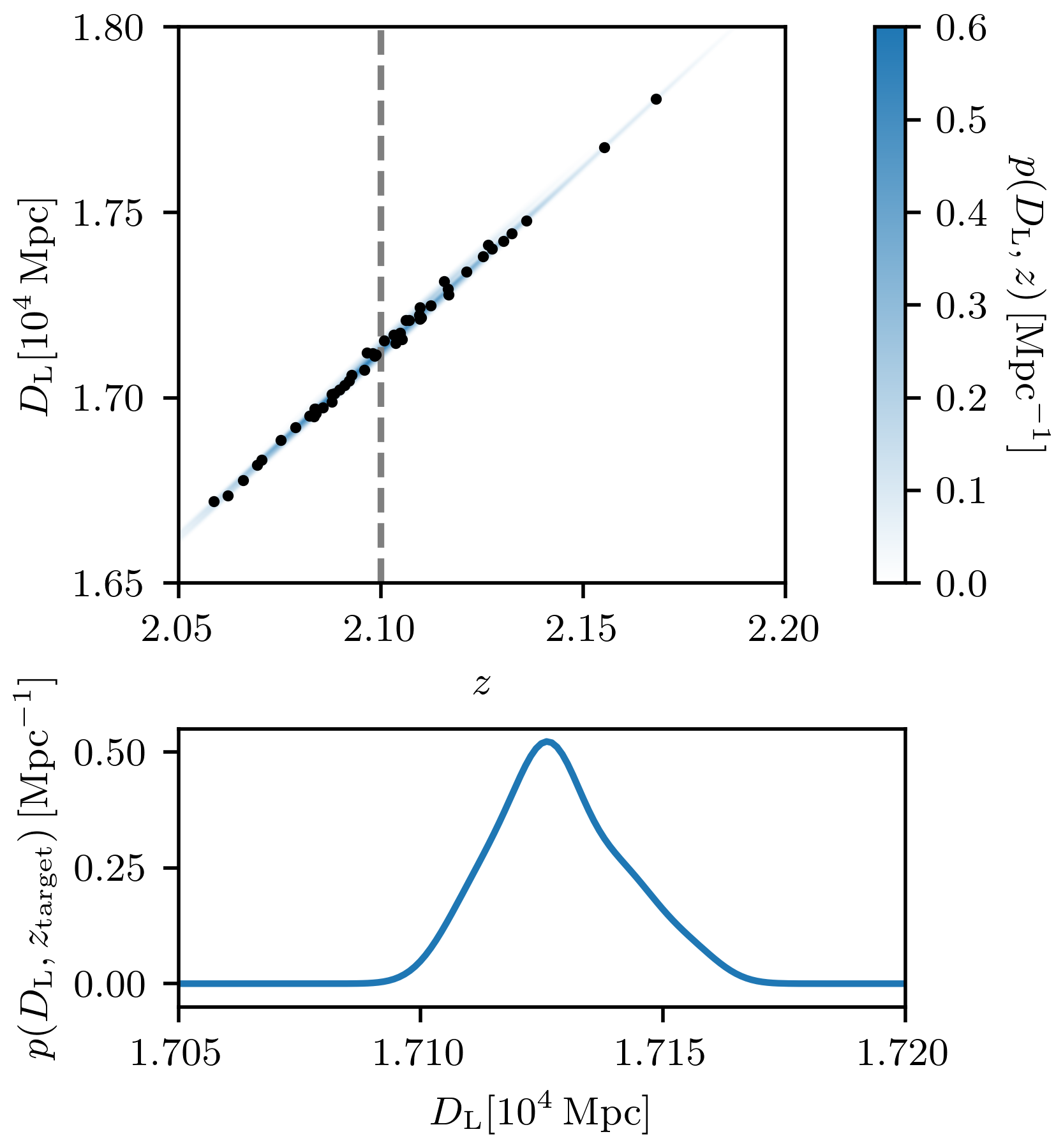}
        \caption{An illustrative example of the $\DL$ dispersion near $\ztarget = 2.1$. \textit{Top panel}. Sample of 50 $(\DL, z)$ pairs in the range $2.05 < z < 2.20$ in a Swiss-cheese universe with standard parameters (see Table \ref{tab:SwissCheeseParams}), $\gamma = 1.1$, $R_{\rm min} = 10$ Mpc, and $R_{\rm max} = 70$ Mpc (black points). The colour gradient displays the kernel density estimate of the normalized PDF $p(\DL, z)$ (see colour bar at right). Although hard to discern by eye, the colour gradient is roughly an ellipse, not a line. The gray dashed line marks $z=\ztarget$. \textit{Bottom panel}. Slice of the PDF $p(\DL, \ztarget)$ calculated by evaluating the kernel density estimate for $z=\ztarget$. Note that $p(\DL, z)$ is normalized on the $\DL$-$z$ plane, but $p(\DL, \ztarget)$ is an unnormalized slice of $p(\DL,z)$.}
        \label{fig:blob}
    \end{figure}

    Fig.\ \ref{fig:blob} plots a sample of 50 $(z, \DL)$ pairs in a Swiss-cheese universe with $\gamma = 1.1$, $R_{\rm min} = 10$ Mpc, $R_{\rm max} = 70$ Mpc, and the standard parameters in Table \ref{tab:SwissCheeseParams}.
    Each pair is generated independently using the above recipe. 
    That is, we do not re-use holes. 
    The pairs display a strong correlation between $\DL$ and $z$, as expected, because the stochastic lensing effect is dominated by the geometric and cosmological expansion of the beam cross-section. 
    The points further away from the grey dashed line correspond to lines of sight, where the final hole is larger. 
    
    To quantify the distribution of the $(\DL, z)$ points around $\ztarget$, we estimate the joint PDF $p(\DL, z)$ by computing a two-dimensional kernel density estimate \citep{WandJones1994, Scott2015} on the 50-point sample plotted in the top panel of Fig.\ \ref{fig:blob}. 
    We adopt a bivariate gaussian kernel.
    The kernel bandwidth is selected using Silverman's rule \citep{Silverman1986}. 
    The resulting density field is plotted as the colour gradient. 
    A slice of $p(\DL, z)$ versus $\DL$ at $z=\ztarget$ is plotted in the bottom panel of Fig.\ \ref{fig:blob}. 
    The PDF slice is not the same as $p(\DL, z)$ marginalised over $z$.
    The peak of $p(\DL, \ztarget)$ occurs at $1.713\times10^{4}\,{\rm Mpc}$, which is systematically higher than $\DL$ at $z=2.1$ in a FLRW universe with matter density $\rho_{\rm cheese}$, spatial curvature $k_{\rm cheese}$, and cosmological constant $\Lambda$ (see Table \ref{tab:SwissCheeseParams}).
    This systematic bias is expected, because the beam propagates mostly through regions which are less dense than $\rho_{\rm cheese}$ \citep{DyerRoeder1972, DyerRoeder1973, Fleury2014}.
    Since the aim of this paper is to quantify the relative contribution to the $\DL$ dispersion of small voids compared to large voids, we do not seek to calibrate precisely the peak of $p(\DL, \ztarget)$ to observed cosmological data.
    The PDF slice is slightly asymmetric about the peak, with a kink at $\DL \approx 1.714\times10^{4}\,{\rm Mpc}$.
    The shape of the PDF slice does not match exactly either the PDF approximated analytically \citep{Wang1999, WangEtAl2002} or the ones approximated by Monte Carlo experiments \citep{BrouzakisEtAl2008, FlanaganEtAl2012}. 
    However, the exact shape is less important in this paper, where we focus on moment-based measures of dispersion, e.g.\ the standard deviation.
    The shape depends on the number of samples and kernel bandwidth and is an approximation of $p(\DL, \ztarget)$.
    We quantify the width of the PDF slice by calculating
    \begin{align}
        \sigma_{\DL} = \qty[\frac{\int \dd \DL' \, \qty(\DL' - \langle\DL\rangle)^2 p(\DL', \ztarget)}{\int \dd \DL' \, p(\DL', \ztarget)}]^{1/2}~,
        \label{eq:sigmaDL_defn}
    \end{align}
    with
    \begin{align}
        \langle\DL\rangle = \frac{\int \dd \DL' \, \DL' p(\DL', \ztarget)}{\int \dd \DL' \, p(\DL', \ztarget)}~.
        \label{eq:avgDL_defn}
    \end{align}    
    The cut-offs of the truncated power-law PDF \eqref{eq:powerlawPDF} ensure that \eqref{eq:sigmaDL_defn} and \eqref{eq:avgDL_defn} do not diverge.
    For the sample shown in the top panel of Fig.\ \ref{fig:blob}, we find $\sigma_{\DL} = 14\,{\rm Mpc}$, which corresponds to $\approx 0.082\%$ dispersion in $\DL$. 
    We compare the latter value with others in the literature in Section \ref{sec:conclusion}.

%% file: section_5.tex
\section{Varying the hole size distribution}
\label{sec:varying_hole_distribution}

To explore the $\DL$ dispersion as a function of $\gamma$ and $R_{\rm min}$, we consider a set of Swiss cheese universes with $-1 \leq \log_{10} (R_{\rm min} / 1\, {\rm Mpc}) \leq 1$ and $1.1 \leq \gamma \leq 3.1$, parameterized on a log-linear grid.
For simplicity, we fix $R_{\rm max} = 70\,{\rm Mpc}$ \citep{HoggEtAl2005, ContariniEtAl2023}.
For each universe, we sample 50 $(z, \DL)$ pairs using the recipe in Section \ref{sec:measuring_DL_dispersion}, at regular intervals of $\ztarget$ in the range $0.5 \leq \ztarget \leq 2.1$, and estimate $p(z, \DL)$ by computing a kernel density estimate. 
We then calculate $\sigma_{\DL}$, the standard deviation of $p(\DL|z=\ztarget)$ defined in Section \ref{sec:measuring_DL_dispersion}.

\begin{figure*}[ht!]
    \centering
    \includegraphics[width=0.98\linewidth]{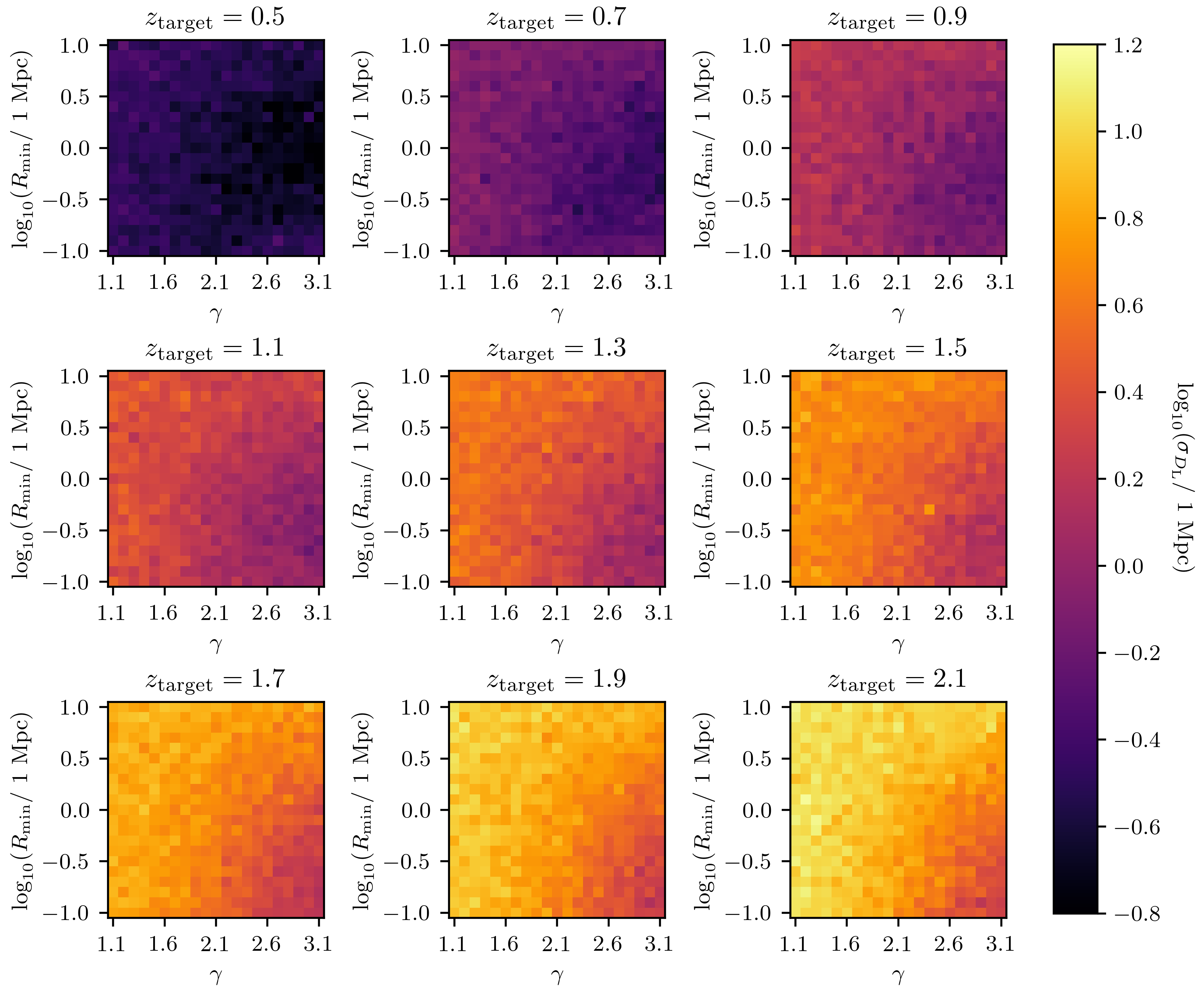}
    \caption{Luminosity distance dispersion as a function of Swiss-cheese hole size distribution. 
    Each panel displays the standard deviation $\sigma_{\DL}$ as a function of the lower cut-off $R_{\rm min}$ (logarithmic vertical axis; $R_{\rm min}$ in units of Mpc) and exponent $\gamma$ (horizontal axis; dimensionless) of the power-law hole size distribution for one redshift in the range $0.5 \leq \ztarget \leq 2.1$ per panel.  
    In each panel, a pixel correspond to a Swiss-cheese universe with standard parameters (see Table \ref{tab:SwissCheeseParams}), $R_{\rm max} = 70\, {\rm Mpc}$, $R_{\rm min}$ in the range $-1 \leq \log_{10} (R_{\rm min} / 1\, {\rm Mpc}) \leq 1$, and $\gamma$ in the range $1.1 \leq \gamma \leq 3.1$. 
    The colour of each pixel displays $\sigma_{\DL}$ in units of Mpc in logarithmic scale (see colour bar at right).
    For each universe, $\sigma_{\DL}$ is calculated using the method detailed in Section \ref{sec:measuring_DL_dispersion} from a sample of 50 $(z, \DL)$ pairs generated independently using the recipe in Section \ref{sec:measuring_DL_dispersion}. 
    For all Swiss-cheese universes, we have $R_{\rm max} = 70\, {\rm Mpc}$, and all other parameters are listed in Table \ref{tab:SwissCheeseParams}.}
    \label{fig:cheeseboard}
\end{figure*}

Figure \ref{fig:cheeseboard} displays $\sigma_{\DL}$ for each universe above at redshifts $0.5 \leq \ztarget \leq 2.1$. 
In every panel, the pixel colour shows $\sigma_{\DL}$ at each $\ztarget$ in each universe. 
Visual inspection reveals that, at fixed $\ztarget$, $\sigma_{\DL}$ decreases smoothly from the top left corner to the lower right corner of the $\gamma$-$R_{\rm min}$ plane. 
Qualitatively, this behavior answers the question we posed in Section \ref{sec:intro}: the $\DL$ dispersion is dominated by a few large voids rather than the many small voids; $\sigma_{\DL}$ decreases towards the lower right corner of each panel in Fig. \ref{fig:cheeseboard}, where $\gamma$ is largest and $R_{\rm min}$ is smallest, corresponding to Swiss-cheese universes with the most small holes.

\begin{figure}[ht!]
    \centering
    \includegraphics[width=0.98\linewidth]{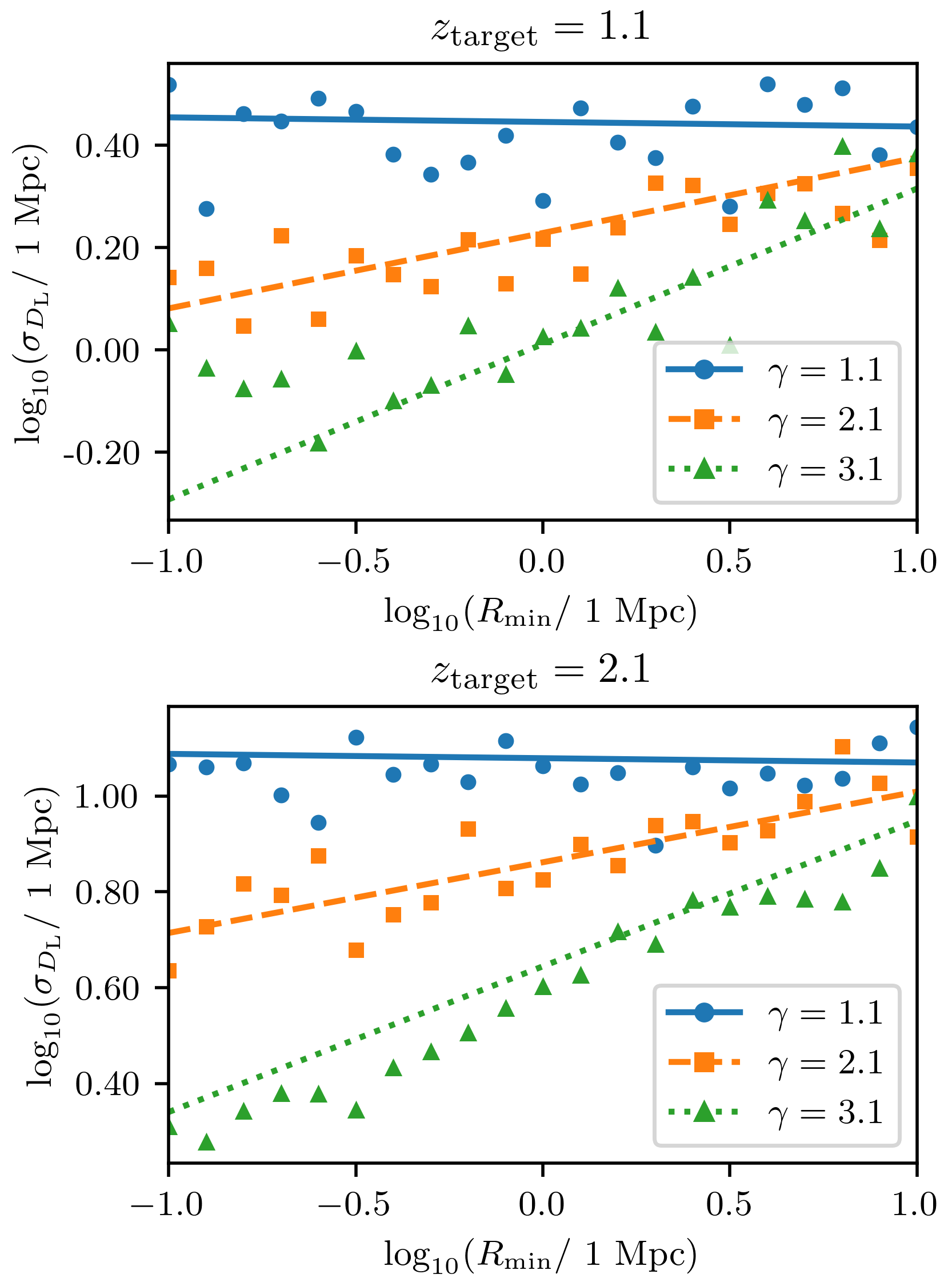}
    \caption{
    Scaling of $\sigma_{\DL}$ as a function of $R_{\rm min}$ and $\gamma$. \textit{Top panel}: $\sigma_{\DL}$ at $\ztarget = 1.1$ for Swiss-cheese universes with $-1 \leq \log_{10} (R_{\rm min} / 1\, {\rm Mpc}) \leq 1$. 
    For each value of $\log_{10} (R_{\rm min} / 1\, {\rm Mpc})$, $\sigma_{\DL}$ is calculated for $\gamma = 1.1, 2.1,$ and $3.1$ (blue circles, orange squares, and green triangles respectively).
    The blue solid line, orange dashed line, and green dotted lines correspond to \eqref{eq:sigma_fit} with the best-fit parameters reported in Section \ref{sec:varying_hole_distribution}.
    \textit{Bottom panel}: as for top panel but with $\ztarget = 2.1$.
    }
    \label{fig:crosssection}
\end{figure}

It is clear from Fig.\ \ref{fig:cheeseboard} that there is a smooth functional relationship between $\sigma_{\DL}$, $\ztarget$, $\gamma$, and $R_{\rm min}$. 
To investigate this relationship, we plot $\sigma_{\DL}$ as a function of $\log_{10} (R_{\rm min} / 1\, {\rm Mpc})$ for three representative $\gamma$ values and two representative $\ztarget$ values in Figure \ref{fig:crosssection}. 
The $\sigma_{\DL}$ values plotted in Fig.\ \ref{fig:crosssection} are the same as the ones plotted in Fig.\ \ref{fig:cheeseboard}. 
We find that, at both representative $\ztarget$ values, $\log_{10} (\sigma_{\DL} / 1\, {\rm Mpc})$ increases approximately linearly with $\log_{10} (R_{\rm min} / 1\, {\rm Mpc})$, and the log-log slope increases with $\gamma$. 
To quantify $\sigma_{\DL}$ as a function of $z$ and the hole size distribution parameters, we perform a nonlinear least-squares fit of the form
\begin{align}
    \sigma_{\DL} = \sigma_0 \ztarget^a \qty(\frac{R_{\rm min}}{R_0})^{b(\gamma - \gamma_0)}
    \label{eq:sigma_fit}
\end{align}
across the whole parameter space (i.e.\ every pixel in every panel of Fig.\ \ref{fig:cheeseboard}).
We obtain the best-fit parameters $\sigma_0 = 2.19\pm0.02 \,{\rm Mpc}$, $a=2.25\pm0.01$, $b=0.157\pm0.003$, $\gamma_0=1.16\pm0.02$, and $R_0 = 24\pm1 \, {\rm Mpc}$.

We graph \eqref{eq:sigma_fit} evaluated with the best-fit parameters in both panels of Fig.\ \ref{fig:crosssection} as blue solid lines, orange dashed lines, and green dotted lines for $\gamma = 1.1, 2.1,$ and $3.1$ respectively. 
In the bottom panel of Fig.\ \ref{fig:crosssection}, the best fit scaling (green dotted line) overestimates $\sigma_{\DL}$ from the numerical experiments (green triangles) systematically. 
The mismatch arises, because the functional form \eqref{eq:sigma_fit} is an empirical approximation.
The approximation breaks down at low $\ztarget$ and low $\gamma$. Nevertheless, we attempt to fit the same functional form across the full domain, which produces a mismatch at high $\ztarget$ and high $\gamma$ as well.
Overall, the best fit scaling \eqref{eq:sigma_fit} matches $\sigma_{\DL}$ from the numerical experiments to within a factor of ${\sim}2$. 

%% file: section_6.tex
\section{Observational and simulation-based hole size distributions}
\label{sec:void_distribution}

\begin{table*}[ht!]
    \centering
    \begin{tabular}{lrrrr}
        \hline
        Void Catalog & Type & Redshift & $L$ ($h^{-1}$ Mpc) & Void size distribution \\
        \hline
        \citet{HoyleVogeley2002}        & Observed & $z < 0.028$ & 20--36 & VPF/UPF \\
        \citet{HoyleVogeley2004}        & Observed & $z < 0.138$ & 10--25 & VPF/UPF \\
        \citet{PanEtAl2012}             & Observed & $z < 0.107$ & 10--30 & -- \\
        \citet{ContariniEtAl2023}       & Observed & $0.2 \leq z \leq 0.65$ & 35--100 & Vdn \\
        \citet{SongEtAl2025b}           & Observed & $0.2 \leq z \leq 0.8$ & 40--113 & Vdn \\
        \hline
        \citet{ContariniEtAl2022}       & Simulated & $0.9 \leq z \leq 1.8$ & 22--50 & Vdn \\
        \citet{WilliamsEtAl2024}        & Simulated & $z \approx 0$ & 8--72 & -- \\
        \hline
    \end{tabular}
    \caption{A selection of void catalogs and their properties. The linear void size $L$ refers to the maximum or effective void radius in each catalog. The acronyms ``VPF'' and ``UPF'' refer to the void probability function and underdensity probability function respectively \citep{White1979, VogeleyEtAl1991}. The acronym ``Vdn'' refers to the volume-conserving void model, which is based on the excursion set formalism \citep{JenningsEtAl2013}.}
    \label{tab:void_catalogs}
\end{table*}

One may ask how the hole size distribution \eqref{eq:powerlawPDF}, with its associated parameter domain in Figure \ref{fig:cheeseboard}, compares to the void size functions observed in galaxy catalogs and cosmological simulations. 
Table \ref{tab:void_catalogs} presents a representative sample of void catalogs in the literature, the redshift range covered, and the detected range of linear void sizes. 

In contemporary void analyses, the comoving number density of voids as a function of their radii is fit typically to the theoretical expression derived using excursion set theory, commonly referred to as the Vdn model \citep{JenningsEtAl2013}.
In the Vdn model, the comoving number density of voids decreases with void size $L$ and is concave down. 
Specifically, the scaling deviates from a power law at low $L$, because small voids are subsumed by larger ones \citep[see, e.g.\ Figure 1 in ][]{JenningsEtAl2013}.
However, the linear sizes of voids in the observed and simulated catalogs typically span less than an order of magnitude, and the minimum linear size of voids is limited by survey resolution to $\gtrsim 10$ Mpc. 
For example, the void counts reported in Figure 3 of \citet{SongEtAl2025b} spans a factor of $\lesssim 3$ in void size.
Hence the linear size below which the comoving number density deviates from a power law is challenging to pinpoint exactly. 
In this paper, we explore the scenario where this scale is smaller than the resolution attainable by observed and simulated galaxy catalogs, by exploring $R_{\rm min}$ across several orders of magnitude below the typical resolution limit.

The limited range of detected void sizes also makes the logarithmic slope of the hole size distribution challenging to determine. 
Visual inspection of Figure 3 in \citet{SongEtAl2025b} shows that the comoving number density of voids scales approximately $\propto L^{-3.6}$ in the reported range. 
The probability of a line of sight intersecting a void of size $L$ is proportional to the product between the comoving number density of voids and their cross-sectional area. 
Hence, the probability that a line of sight will encounter a void of size $L$ is proportional to $L^{-1.6}$ for the representative \citet{SongEtAl2025b} scaling, corresponding to $\gamma = 1.6$.
A similar argument can be made based on other observed and simulation-based void size distributions included in Table \ref{tab:void_catalogs}. 
The value of $\gamma$ varies from catalog to catalog, with $1 \lesssim \gamma \lesssim 4$.
The wide range of plausible $\gamma$ is a result of the limited size range of detectable voids.
The numerical experiments in Section \ref{sec:varying_hole_distribution} is computationally expensive for larger $\gamma$, because the number of holes that the light beam passes through increases by orders of magnitude. 

%% file: section_7.tex
\section{Extending to higher redshift}
\label{sec:highz}

The analysis in Section \ref{sec:varying_hole_distribution} focuses on the redshift range $0.5 \lesssim z \lesssim 2.1$ relevant to SNe Ia experiments. 
Other studies of the void distribution, such as gravitational wave observations of compact binary coalescences (see below) \citep{VitaleWhittle2018, KalogeraEtAl2021}, are likely to probe higher $z$ in the future. 
In preparation for such studies and as a proof of principle, we extend the numerical calculations of $\sigma_{\DL}$ to a representative range of higher redshifts in this section.

We consider the same set of Swiss-cheese universes as in Section \ref{sec:varying_hole_distribution}. 
For each universe, we calculate $\sigma_{\DL}$ at $\ztarget = 2.0, 3.0, 4.0,$ and $5.0$ using the procedure described in Section \ref{sec:measuring_DL_dispersion}. We compare the $\sigma_{\DL}$ calculated numerically with the $\sigma_{\DL}$ calculated by extrapolating the empirical scaling \eqref{eq:sigma_fit}.
We find that the extrapolated scaling overestimates $\sigma_{\DL}$ typically.
This mismatch worsens as redshift increases, reaching a factor $\lesssim 2$ at $z=5.0$.
The empirical scaling of the form \eqref{eq:sigma_fit} does not fit well across the whole parameter domain.

\begin{figure}[ht!]
    \centering
    \includegraphics[width=0.98\linewidth]{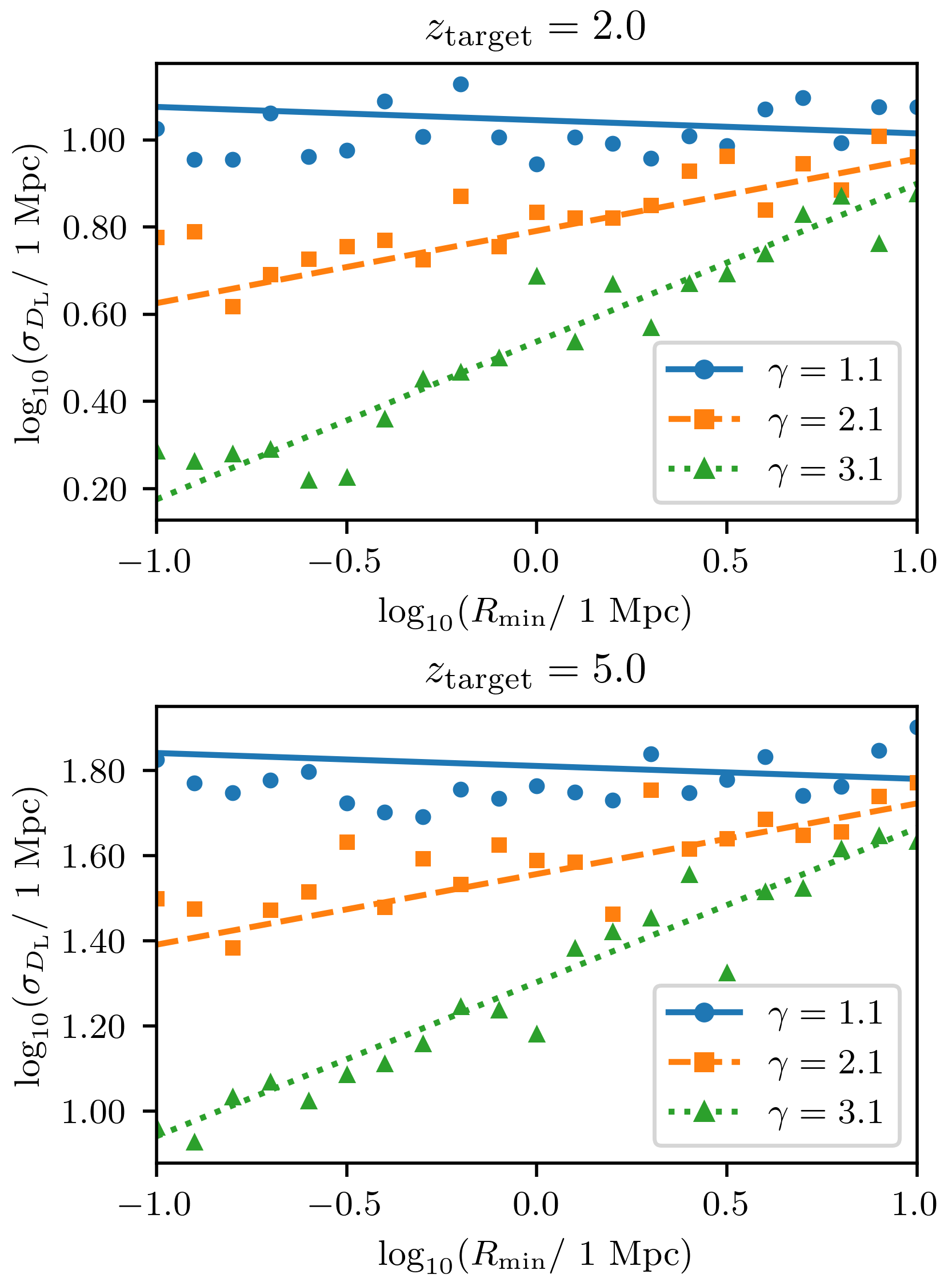}
    \caption{
    As for Figure \ref{fig:crosssection}, but with $z_{\rm target}=2.0$ (\textit{top panel}), and $z_{\rm target}=5.0$ (\textit{bottom panel}), with the best-fit parameters reported in Section \ref{sec:highz}.}
    \label{fig:highz}
\end{figure}

Instead of attempting an exhaustive trial and error to find a functional form that fits well to all $\ztarget$, we perform a nonlinear least-squares fit of equation \eqref{eq:sigma_fit} to the data with $\ztarget = 2.0, 3.0, 4.0,$ and $5.0$.  
We obtain the best-fit parameters $\sigma_0 = 2.67\pm0.05 \,{\rm Mpc}$, $a=1.92\pm0.01$, $b=0.196\pm0.004$, $\gamma_0=1.25\pm0.02$, and $R_0 = 19\pm1 \, {\rm Mpc}$.
These revised, high-$z$ parameters complement the low-$z$ parameters quoted in the penultimate paragraph of Section \ref{sec:varying_hole_distribution}.
Figure \ref{fig:highz} displays $\sigma_{\DL}$ for a subset of these universes at three representative $\gamma$ values, $\gamma=1.1, 2.1,$ and $3.1$, at two representative $\ztarget$ values, $\ztarget = 2.0$ (minimum) and $5.0$ (maximum).
The scaling \eqref{eq:sigma_fit} evaluated with the high-$z$ best-fit parameters for $\gamma=1.1, 2.1,$ and $3.1$ are plotted in the top and bottom panels of Fig.\ \ref{fig:highz} as blue solid lines, orange dashed lines, and green dotted lines respectively.
The best-fit scaling matches $\sigma_{\DL}$ from the numerical experiments to within $\approx 20\%$.

Third-generation (3G) gravitational wave observatories are expected to detect gravitational radiation from compact binary coalescences for $z \gg 1$ \citep{VitaleWhittle2018, KalogeraEtAl2021}.
The typical $\sigma_{\DL}$ at $z \sim 5$ displayed in Fig.\ \ref{fig:highz} ($\sim 60$ Mpc) is comparable to the expected luminosity distance uncertainty at design sensitivity of 3G detectors assuming standard $\Lambda$CDM cosmology \citep[$\sim 30$ Mpc at $z\sim 5$; see Fig. 5 in][]{VitaleWhittle2018}. 
Stochastic gravitational lensing by cosmic voids may therefore be an important source of uncertainty for Hubble constant measurements using standard sirens for 3G observatories \citep{Schutz1986, AbbottEtAl2017, AbbottEtAl2023}.
It may be possible to reduce the $\DL$ dispersion due to stochastic lensing by forward-modelling the foreground matter distribution \citep{AmendolaEtAl2010, KronborgEtAl2010, JonssonEtAl2010, SmithEtAl2013, ShahEtAl2024}.

%% file: section_8.tex
\section{Conclusion}
\label{sec:conclusion}

The luminosity distance of a standard candle depends in part on the stochastic distribution of intervening spacetime curvature between the source and the observer, and therefore differs from one line of sight to the next.
In this paper, we investigate contributions to the $\DL$ dispersion due to hypothetical small-scale (${\lesssim} 10$ Mpc) inhomogeneities unresolved by galaxy surveys and numerical simulations. 
Specifically, by tracing ray bundles through an ensemble of Swiss-cheese universes with a truncated power-law hole size distribution, we quantify how the $\DL$ dispersion at redshifts $0.5 \leq z \leq 2.1$ scales with the lower cut-off $R_{\rm min}$ and logarithmic slope $\gamma$ of the hole size distribution.
We find that the standard deviation $\sigma_{\DL}$ of $\DL$ is lowest in universes, where lines of sight pass through the most small holes.
A nonlinear least-squares fit to simulated $(\DL, z)$ data yields the scaling $\sigma_{\DL} \propto z^a R_{\rm min}^{b(\gamma - \gamma_0)}$, with $a=2.25\pm0.01$, $b=0.157\pm0.003$, and $\gamma_0=1.16\pm0.02$, according to equation \eqref{eq:sigma_fit}.

The dispersion $\sigma_{\DL}$ is an order of magnitude smaller than the one estimated by \citet{FlanaganEtAl2012} and \citet{BrouzakisEtAl2008} for Swiss-cheese universes with hole size $\gtrsim 30\,{\rm Mpc}$.  
This is expected, because the density profiles used in the foregoing references contain denser mass-compensating regions, causing a larger lensing effect. 
For example, the density profile displayed in fig.\ 1 of \citet{BrouzakisEtAl2008} reaches $\rho(t, r)/\rho_{\rm cheese} \lesssim 2.5$, while the Emmentaler profile maintains $\rho(t, r)/\rho_{\rm cheese} \lesssim 1.2$ throughout its evolution.

The sub-Mpc voids considered in this paper are hypothetical. 
The distribution of voids plausibly flattens out or cuts off towards the low-$L$ end, as predicted by the excursion set formalism \citep{JenningsEtAl2013}. 
The result displayed in Fig.\ \ref{fig:cheeseboard} indicates that even if small voids are abundant, their contribution to the $\DL$ dispersion would be subdominant.
On the other hand, astrophysical voids differ not only by their size, but also by features such as their central underdensity and the shape of their density profile \citep{HamausEtAl2014, WilliamsEtAl2024}.
The dispersion in these features also contributes to $\sigma_{\DL}$ in general.
Quantifying these contributions systematically lies outside the scope of this paper. 

The $\DL$ dispersion computed in this paper for an ensemble of universes is prompted by the absence of information about the inhomogeneous distribution of matter along lines of sight in the actual universe. 
If instead the foreground matter can be mapped with sufficient precision, and modeled with sufficient detail, one can avoid treating the dispersion probabilistically. 
For instance, \citet{ShahEtAl2024} demonstrated that forward-modelling the weak lensing effect by the extragalactic foreground reduces the $\DL$ dispersion of the SNe at $z\sim1$ in the Dark Energy Survey SN Ia Year 5 sample by ${\sim}10$ per cent.

%% file: thanks.tex
\newpage
\section*{Acknowledgements}
The authors thank Kok Hong Thong, Wenhao Dong, and Tamara Davis for helpful discussions. 
We also thank Nicholas Kah Yean Low for assistance with numerical computation. 
We acknowledge support from the Australian Research Council (ARC) through the Centre of Excellence for Gravitational Wave Discovery (OzGrav) (grant numbers CE170100004 and CE230100016).
TC is supported by an Australian Government Research Training Program Scholarship. 
This research was supported by The University of Melbourne's Research Computing Services and the Petascale Campus Initiative. 

%% file: appendix_a.tex
\section{Israel junction conditions}
\label{appx:junctioncondition}

In this appendix, we write down the Israel conditions for matching the LTB metric in a hole to the FLRW metric in the cheese. 
We assume in conjunction with the main text that there is no thin mass shell on the boundary between any hole and the surrounding cheese. 
We choose spherical polar coordinates $(t, r, \theta, \phi)$ such that the $t$-coordinate is the time measured by comoving observers, and the $r$-coordinate is continuous from hole to cheese, with the region $r < R$ being the hole, and $r > R$ being the cheese. 
Using this coordinate system, the hypersurface on which the Israel conditions apply is given by $r = R$.

The first Israel condition for joining two metrics on a hypersurface $\Sigma$ is 
\begin{align}
    \qty[h_{ab}] = 0~, \label{eq:Israel_first_general}
\end{align}
where $h_{ab}$ is the induced metric on $\Sigma$ \citep{Israel1966, Poisson2004}. 
In \eqref{eq:Israel_first_general}, square brackets denote the jump of the enclosed expression across $\Sigma$.
Using the Friedmann-like parameterization of the LTB metric in Section \ref{subsec:friedmannlikeparam}, we evaluate equation \eqref{eq:Israel_first_general} on the hypersurface $r=R$ and obtain
\begin{align}
    a(t, R) = a_{\rm cheese}(t) \label{eq:Israel_first}
\end{align} 
for all $t$. 
Specifically, equation \eqref{eq:Israel_first} evaluates to $a(t_{\rm now}, R) = a_{\rm cheese}(t_{\rm now})$.

The second Israel condition is
\begin{align}
    S_{ab} = - \frac{\epsilon}{8\pi} \qty[K_{ab} - K h_{ab}]~, \label{eq:Israel_second_general}
\end{align}
where $S_{ab}$ denotes the surface stress-energy tensor of a thin mass shell on $\Sigma$, $K_{ab}$ denotes the extrinsic curvature of $\Sigma$, and we write $K=h^{ab}K_{ab}$ \citep{Israel1966, Poisson2004}. 
Again, in \eqref{eq:Israel_second_general}, square brackets denote the jump of the enclosed expression across $\Sigma$.
We assume in conjunction with Section \ref{subsec:junction-compensation} that there is no thin mass shell surrounding the holes, viz.\ $S_{ab} = 0$. 
With the Friedmann-like parameterization of the LTB metric, equation \eqref{eq:Israel_second_general} reduces to
\begin{align}
    R a(t, R) \qty[ 1 - k(R)R^2 ] = R a_{\rm cheese}(t) \qty[ 1 - k_{\rm cheese}R^2 ] \label{eq:Israel_second}
\end{align} 
on the hypersurface $r=R$.
Substituting \eqref{eq:Israel_first} into \eqref{eq:Israel_second} gives 
\begin{align}
    k(R) = k_{\rm cheese}~. \label{eq:Israel_k}
\end{align} 

To guarantee that \eqref{eq:Israel_first} is satisfied for all $t$, we take the $t$-derivative of \eqref{eq:Israel_first} and substitute equations \eqref{eq:FriedmannLikeEoM}, \eqref{eq:Israel_k}, and the Friedmann equation for $a_{\rm cheese}(t)$. 
The result is
\begin{align}
    \mu(R) = 4\pi G \rho_{\rm cheese} / 3,
\end{align}
where $\rho_{\rm cheese}$ is the matter density in the FLRW cheese.


%% file: appendix_b.tex
\section{Integrating the equations of motion}
\label{appx:metricfunctions}

Including \eqref{eq:da_ds} and \eqref{eq:dk_ds} in the equations of motion allows us to avoid solving equation \eqref{eq:integratedFriedmannLikeEoM} for $a[x^t(s), x^r(s)]$ and $k[x^r(s)]$ at each $s$-step. 
However, the right-hand sides of equations \eqref{eq:dpt_ds}--\eqref{eq:dxphi_ds}, and  \eqref{eq:db_ds}--\eqref{eq:dk_ds} depend not only on $a[x^t(s), x^r(s)]$ and $k[x^r(s)]$ but also their derivatives $\dot{a}, \dot{a}', a', a'', k'$. 
In this appendix, we discuss how these derivatives can be calculated using only quantities in the beam state vector $\vb{V} = \qty(p^t, p^r, x^t, x^r, x^\phi, b, J, a, k)$. 

First, we note that equation \eqref{eq:FriedmannLikeEoM} determines $\dot{a}$ in terms of $a$ and $k$. 
To calculate the remaining derivatives, we start by integrating equation \eqref{eq:FriedmannLikeEoM} from $t$ to $t_{\rm now}$. 
The result is
\begin{equation}
    \int_{a(t, r)}^{1} \dd \tilde{a} \; \qty[\frac{2\mu(r)}{\tilde{a}} - k(r) + \frac{\Lambda \tilde{a}^2}{3}]^{-1/2} = t_{\rm now} - t ~. \label{eq:integratedFriedmannLikeEoM_appx}
\end{equation}
Taking the derivative of \eqref{eq:integratedFriedmannLikeEoM_appx} with respect to $r$, we obtain
\begin{align}
    a'(t, r) &= -\frac{1}{2}\mathcal{F}(r, k, a)^{1/2} \qty[ - k' \mathcal{U}(r, k, a) + 2\mu'\mathcal{V}(r, k, a)]~, \label{eq:aprime}
\end{align}
with
\begin{align}
    \mathcal{F}(r, k, a) = \frac{\Lambda a^2}{3} - k + \frac{2\mu(r)}{a}~, 
\end{align}
\begin{align}
    \mathcal{U}(r, k, a) = \int_a^1 \dd \tilde{a}~  \mathcal{F}(r, k, \tilde{a})^{-3/2} \label{eq:U}~, 
\end{align}
and 
\begin{align}
    \mathcal{V}(r, k, a) = \int_a^1 \dd \tilde{a}~ \tilde{a}^{-1} \mathcal{F}(r, k, \tilde{a})^{-3/2} \label{eq:V}~. 
\end{align}
We suppress the arguments of $\mu(r)$, $k(r)$, $a(t, r)$ and their derivatives in \eqref{eq:aprime} for visual clarity. 
Taking another derivative of \eqref{eq:aprime} with respect to $r$ yields
\begin{align}
    a''(t, r) &= a'\mathcal{F}(r, k, a)^{-1} \qty(\frac{\Lambda aa'}{3} - k' + \frac{2\mu'}{a} - \frac{\mu a'}{a^2}) \nonumber \\
            &\phantom{=}+\frac{1}{2}\mathcal{F}(r, k, a)^{1/2}  
            \left[k''\mathcal{U}(r, k, a) - 2 \mu''\mathcal{V}(r, k, a) + \frac{3}{2}\mathcal{W}(r, k, a) \right] ~, \label{eq:aprimeprime}
\end{align}
with
\begin{align}
    \mathcal{W}(r, k, a) = \int_{a}^1 \dd \tilde{a} ~ \qty(\frac{2 \mu'}{\tilde{a}}-k')^2  \mathcal{F}(r, k, \tilde{a})^{-5/2}~. \label{eq:W}
\end{align}

Evaluating \eqref{eq:aprime} and \eqref{eq:aprimeprime} at $t=t_{\rm B}(r)$, where we have $a[t_{\rm B}(r), r] = 0$, and rearranging gives
\begin{align}
    k'(r) &= 2\mu'(r) \mathcal{V}(r, k, 0) / \mathcal{U}(r, k, 0) ~, \label{eq:kprime}
\end{align}
and 
\begin{align}
    k''(r) &=  \frac{2\mu'' \mathcal{V}(r, k, 0)}{\mathcal{U}(r, k, 0)} - \frac{3\mathcal{W}(r, k, 0)}{2\mathcal{U}(r, k, 0)} ~. \label{eq:kprimeprime}
\end{align}
Lastly, taking the derivative of \eqref{eq:FriedmannLikeEoM} with respect to $r$, we obtain
\begin{align}
    \dot{a}'(t, r) &= \qty(\frac{\Lambda aa'}{3} - \frac{k'}{2} + \frac{\mu'}{a} - \frac{\mu a'}{a^2}) \mathcal{F}(r, k, a)^{-1/2}~. \label{eq:adotprime}
\end{align}

At each $s$-step, we calculate $\mathcal{U}(r, k, 0)$, $\mathcal{V}(r, k, 0)$, $\mathcal{W}(r, k, 0)$, $\mathcal{U}(r, k, a)$, $\mathcal{V}(r, k, a)$, and $\mathcal{W}(r, k, a)$ by computing numerically the integrals \eqref{eq:U}, \eqref{eq:V}, and \eqref{eq:W}. 
Then, we calculate $a', a'', k', k''$, and $\dot{a}'$ by evaluating equations \eqref{eq:aprime}, \eqref{eq:aprimeprime}, and \eqref{eq:kprime}--\eqref{eq:adotprime}. 
These evaluations involve only algebraic operations and are cheap computationally. 

In this paper, the definite integrals in \eqref{eq:U}, \eqref{eq:V}, and \eqref{eq:W} are computed by quadrature with the algorithm implemented in the Fortran library QUADPACK \citep{PiessensEtAl1983}, accessed through the Python package \textsc{scipy}. 
Within our computational set-up, the acceleration gained from not solving \eqref{eq:FriedmannLikeEoM} iteratively at every $s$-step outweighs the cost of computing numerically the definite integrals \eqref{eq:U}, \eqref{eq:V}, and \eqref{eq:W}.



%% file: appendix_c.tex
\section{Continuity conditions}
\label{appx:continuity}

In this appendix, we write down the conditions for transforming the the components of a state vector $\vb{V}$ of a light beam exiting one Swiss-cheese hole to the components of $\vb{V}$ entering the next Swiss-cheese hole. 

Consider two adjacent Swiss-cheese holes $\mathcal{H}_{i}$ and $\mathcal{H}_{i+1}$ depicted in Fig.\ \ref{fig:continuity}. 
The beam exits $\mathcal{H}_{i}$ and enter $\mathcal{H}_{i+1}$ at point $\mathcal{O}$, where we place a fictitious observer. 
Let us denote the coordinate system of hole $\mathcal{H}_i$ by $\mathcal{C}_i = (t_i, r_i, \theta_i, \phi_i)$, and the components of $\vb{V}$ at $\mathcal{O}$ in $\mathcal{C}_{i}$ by $\qty(p^{t_i}, p^{r_i}, x^{t_i}, x^{r_i}, x^{\phi_i}, k_i, a_i, b_i, J_i)$. 

The components of $\vb{V}$ in $\mathcal{C}_i$ and in $\mathcal{C}_{i+1}$ are related as follows:
\begin{itemize}
    \item We orient $\mathcal{C}_{i+1}$ to give $x^{\phi_{i+1}} = 0$.

    \item Both $x^{t_i}$ and $x^{t_{i+1}}$ are the proper time of the comoving observer at $\mathcal{O}$, so we have $x^{t_{i+1}} = x^{t_i}$.

    \item By construction, $x^{r_{i+1}}$ is the radius of $\mathcal{H}_{i+1}$. 

    \item Setting $c$ to be equal for every hole leads to $\rho(t, R)/\rho_{\rm cheese}(t)$, and consequently $a(t, R)$ and $k(R)$, being equal for every hole. Therefore, we have $a_{i+1} = a_i$, and $k_i =  k_{i+1}$.

    \item Let us denote the energy of a photon in the beam observed by the observer at $\mathcal{O}$ as $\mathcal{E}$. 
    In $\mathcal{C}_{i}$, we have $\mathcal{E} = p^{t_{i}}$.
    In $\mathcal{C}_{i+1}$, we have $\mathcal{E} = p^{t_{i+1}}$. Therefore, $p^{t_{i+1}} = p^{t_i}$ holds.

    \item The angle $\alpha_{i+1}$ between the beam direction and $\vb{e}_{r_{i+1}}$, the basis vector in the $r_{i+1}$-direction at $\mathcal{O}$, is given by 
    \begin{equation}
        \cos{\alpha_{i+1}} = \frac{\vb{p}\cdot\vb{e}_{r_{i+1}}}{\mathcal{E}(\vb{e}_{r_{i+1}} \cdot \vb{e}_{r_{i+1}})^{1/2}}~. \label{eq:alpha}
    \end{equation}
    Evaluating \eqref{eq:alpha} in $\mathcal{C}_{i+1}$ and rearranging yields
    \begin{align}
        p^{r_{i+1}} &=  p^{t_{i+1}} \cos{\alpha_{i+1}}\qty[1 - k_{i+1} \qty(x^{r_{i+1}})^2]^{1/2} \qty[a_{i+1} + x^{r_{i+1}} a'_{i+1}]^{-1}~,
    \end{align}
    with $a'_{i+1} = a'\qty(x^{t_{i+1}}, x^{r_{i+1}})$.
    The incident angle $\alpha_{i+1}$ is drawn from a distribution as described in step (ii) of the recipe in Section \ref{sec:measuring_DL_dispersion}.
    We calculate $a'_{i+1}$ by substituting $x^{r_{i+1}}$, $k_{i+1}$, and $a_{i+1}$ into \eqref{eq:aprime}. 
    
    \item The cross-sectional area of the beam is coordinate-independent \citep{MisnerEtAl1973}, implying $b_i = b_{i+1}$. 
    
    \item We set the scale of the affine parameter to be continuous along the beam centroid, implying $J_i = J_{i+1}$.
    
\end{itemize}